\documentclass[final,5p,times,twocolumn]{elsarticle}

\usepackage{amssymb}
\usepackage{amsthm}
\usepackage{amsmath}
\usepackage{enumitem}
\usepackage{xcolor}
\usepackage{soul}
\biboptions{sort&compress}
\usepackage{lineno}
\usepackage{tabularx}
\usepackage{multirow}
\usepackage{hyperref}

\journal{Nuclear Instruments and Methods in Physics Research Section A}

\begin{document}
\begin{frontmatter}



\title{Spectral distribution and flux of $\gamma$-ray beams produced through Compton scattering of unsynchronized laser and electron beams}

\author[ifin]{Dan Filipescu}\ead{dan.filipescu@nipne.ro}
\author[ifin]{Ioana Gheorghe}       
\author[msu]{Konstantin Stopani}     
\author[msuff]{Sergey Belyshev}     
\author[lasti]{Satoshi Hashimoto}   
\author[lasti]{\\Shuji Miyamoto}    
\author[konan]{Hiroaki Utsunomiya}  

\address[ifin]{Horia Hulubei National Institute of Physics and Nuclear Engineering -- IFIN-HH, 077125 Bucharest, Romania}
\address[msu]{Lomonosov Moscow State University, Skobeltsyn Institute of Nuclear Physics, 119991 Moscow, Russia}
\address[msuff]{Lomonosov Moscow State University, Faculty of Physics, 119991 Moscow, Russia}
\address[lasti]{Laboratory of Advanced Science and Technology for Industry, University of Hyogo, 3-1-2 Kouto, Kamigori, Ako-gun, Hyogo 678-1205, Japan}
\address[konan]{Konan University, Department of Physics, 8-9-1 Okamoto, Higashinada, Japan}

\begin{abstract}
Intense, quasi-monochromatic, polarized $\gamma$-ray beams with high and tunable energy produced by Compton scattering of laser photons against relativistic electrons are used for fundamental studies and applications. 
Following a series of photoneutron cross section measurements in the Giant Dipole Resonance (GDR) energy region performed at the NewSUBARU synchrotron radiation facility, we have developed the \texttt{eliLaBr} Monte Carlo simulation code for characterization of the scattered $\gamma$-ray photon beams. 
The code is implemented using \textsc{Geant4} and is available on the GitHub repository (https://github.com/dan-mihai-filipescu/eliLaBr).
Here we report the validation of the \texttt{eliLaBr} code on NewSUBARU LCS $\gamma$-ray beam flux and spectral distribution data and two applications performed with it for asymmetric transverse emittance profiles electron beams, characteristic for synchrotrons. 

The first application is based on a systematic investigation of transverse collimator offsets relative to the laser and electron beam axis. We show that the maximum energy of the LCS $\gamma$-ray beam is altered by vertical collimator offsets, where the edge shifts towards lower energies with the increase in the offset.

Secondly, using the \texttt{eliLaBr} code, we investigate the effect of the laser polarization plane orientation on the properties of the LCS $\gamma$-ray beams produced with asymmetric emittance electron beams. We show that:

1. The use of vertically polarized lasers contributes to the preservation of the LCS $\gamma$-ray beam maximum energy edge by increasing the precision in the vertical collimator alignment.

2. Under identical conditions for the electron and laser beams phase-space distributions, the energy spectrum of the scattered LCS $\gamma$-ray beam changes with the laser beam polarization plane orientation. More precisely, the use of vertically polarized laser beams slightly deteriorates the LCS $\gamma$-ray beam energy resolution.

\end{abstract}


\end{frontmatter}

\section{Introduction}
\label{label_sec_intro}

Intense, quasi-monochromatic, polarized $\gamma$-ray beams with high and tunable energy produced by Compton scattering of laser photons (LCS) against relativistic electrons find applications in nuclear physics \cite{Shizuma2021,KEIde2021,Weller2009}, medical radioisotopes production \cite{WLuo2016}, MeV $\gamma$-ray astronomy \cite{PGros2018}, non-destructive inspection techniques \cite{HLan2021} and electron beam diagnostics \cite{Klein2002,Sun2009_STAB,Chouffani2006}.
Today, LCS $\gamma$-ray beams are produced at several research infrastructures, such as HI$\gamma$S~\cite{litvinenko_1997} in the USA, NewSUBARU~\cite{amano09,Horikawa2010} and LEPS2~\cite{muramatsu2022} at SPring8~--~Japan, SLEGS~\cite{wang_fan_2022} in China, with various new facilities currently in plan \cite{ZPan2019,Micieli2016,facetII,elinp_web}  or under construction \cite{HZen2016}. 

The spectral distribution and flux of LCS $\gamma$-ray beams in terms of electron beam parameters, laser~--~electron beam collision angle and collimation settings have been extensively discussed in the literature \cite{Klein2002,Angelo2000,Petrillo2012,krafft2016}. The focus has been set on fixed interaction point sources based on synchronized laser and electron beam bunches, where, besides the general beam-beam interaction \textsc{cain} code \cite{CAIN}, we mention the dedicated LCS Monte Carlo simulation codes \textsc{mccmpt}~\cite{Sun2011_STAB} validated on HI$\gamma$S experimental data, the \textsc{mclcss} code \cite{Luo2011} developed for characterization of the SLEGS source, the \textsc{cmcc} code~\cite{Curatolo_PhD_thesis} applied in Ref.~\cite{Curatolo2017} for characterization of the STAR-I, EuroGammaS and XFELO $\gamma$-sources. Recently, the collimation optimization for production of LCS $\gamma$-ray beams through head-on collisions between low energy CO$_2$ photons against NewSUBARU electron beams has been discussed in Ref.~\cite{Hajima2021} based on Monte Carlo simulations. The algorithm given in Ref.~\cite{Hajima2021} has been afterwards updated to account also for arbitrary laser~--~electron angle collisions, produce approximate evaluations of the scattered photon polarization and provide flux information for the LCS $\gamma$-ray beam~\cite{paterno_2022}. 

In this work, we introduce a Monte Carlo simulation method for characterizing the photon flux and spectral distribution of LCS $\gamma$-ray beams. The method has been originally developed for LCS beam diagnostics for photoneutron experiments at the NewSUBARU synchrotron radiation facility. At NewSUBARU, the laser-electron interactions can take place along the entire length of the straight electron beamline BL01, following the spatial overlap between the laser beam and the electron beam along the interaction region. However, in the majority of the existing LCS simulation codes \cite{Sun2011_STAB,Curatolo2017,Hajima2021,paterno_2022}, it is considered that the incident beams collide so that the centers of the laser and electron beam pulses overlap at the designed interaction point, without any position and timing jitters. Thus, in order to model the NewSUBARU LCS source, we found it necessary to implement a Monte Carlo simulation code for Compton interaction between continuous, unsynchronized laser and electron beams sent head-on against each other, with realistic modeling of the spatial overlap between the laser and electron beams along the interaction region.

Besides the main difference mentioned above, between the present and the existing LCS simulation codes, we list the following advantages of the present implementation:

\paragraph{(A) Compatibility with various LCS $\gamma$-ray sources} 
The code can be easily adapted to other facilities, besides the NewSUBARU one, as the following parameters are read from input files:
\begin{itemize}
\item electron beam energy, energy spread and transverse emittance values, where a Gaussian emittance profile is considered. Also the $\alpha$ and $\beta$ Twiss parameters, the dispersion function $D$ and the phase space advance $\psi$, which are used for the determination of the electron beam phase space distribution along the interaction region.
\item laser beam wavelength, energy spread, polarization state, diameter, quality factor $M^2$, as well as optical parameters of the focusing lens. 
\item collimation system parameters: distance to the center of the electron beamline, thickness and aperture.
\end{itemize}

\paragraph{(B) Full modeling of the polarization state for the scattered photon} 
The polarization is treated both within the Stokes parameters formalism and in the vectorial formalism~\cite{Filipescu_2022_POL}. The code provides the polarization properties of the scattered photon in the particle reference system, according to the specific requirements of the \textsc{Geant4} framework.

\paragraph{(C) Integrated into end-to-end experiment simulation tool} 
The LCS $\gamma$-ray source model was implemented in a \texttt{C++} class integrated into the \textsc{Geant4} framework~\cite{geant_agostinelli_2003,geant_allison_2006,geant_allison_2016} to produce realistic end-to-end simulations of photonuclear experiments. A flexible design moderated neutron counter array and LCS $\gamma$-ray flux and spectral distribution monitor detectors of flexible type (material) and dimensions are also implemented in the code.

\paragraph{(D) Open source}
The Monte Carlo simulation code \texttt{eliLaBr} implemented following the present algorithm is available on the GitHub repository~\cite{eliLaBr_github}. 

\begin{figure}[t]
\centering
\includegraphics[width=0.45\textwidth, angle=0]{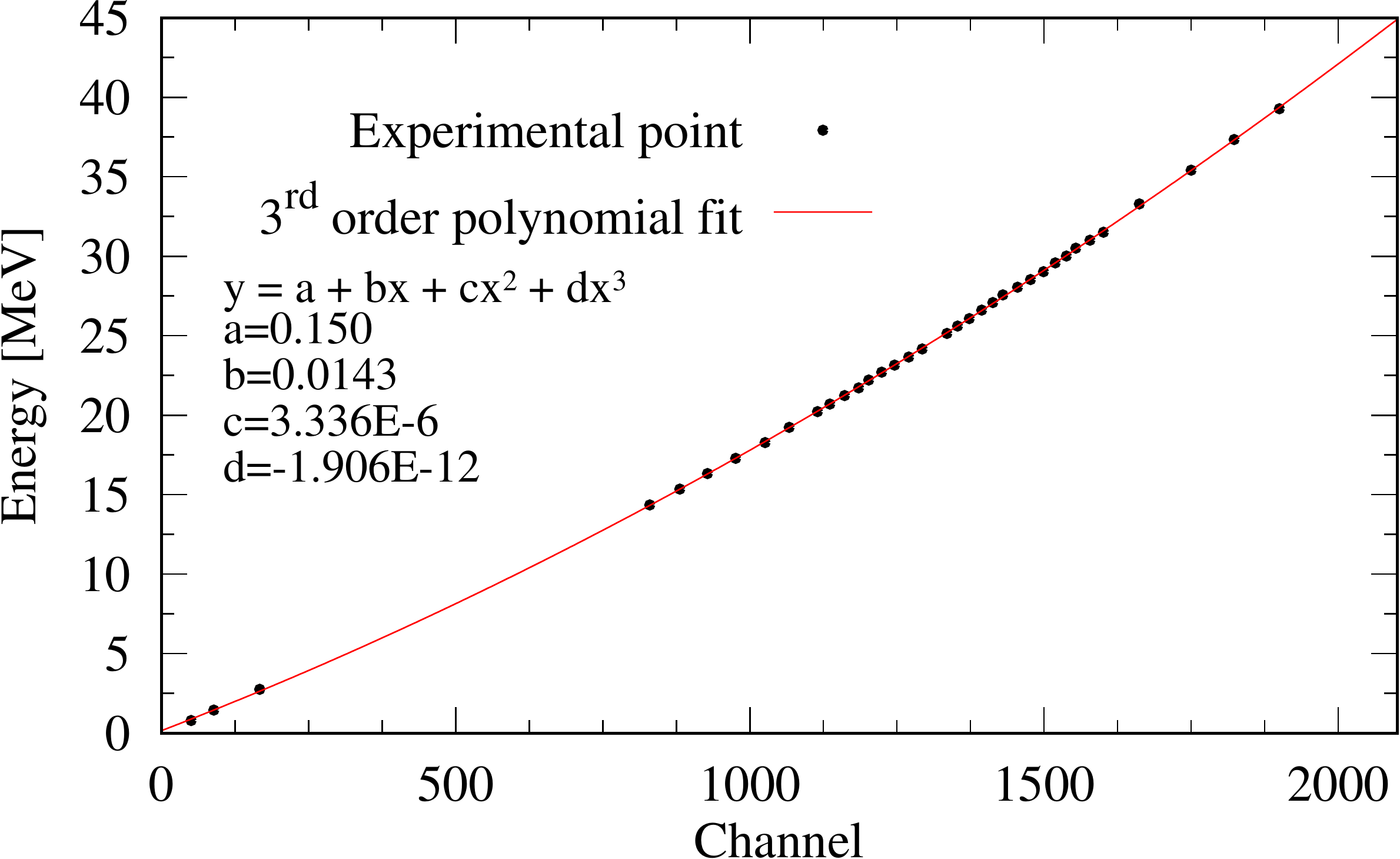} 
\caption{Energy calibration of the 3.5"~$\times$~4.0" LaBr$_3$ detector with internal radiation and the maximum energies of LCS $\gamma$-ray beams produced with a 0.532~$\mu$m laser and electron beams at energies between 641 and 1069~MeV.   
}
\label{fig_labr_cal}       
\end{figure}

In Sec.~\ref{label_sec_motivation} we discuss the motivation of our work: the characterization of high energy LCS $\gamma$-ray beam spectra for accurate photoneutron cross section measurements in and above the GDR region. 
Section~\ref{label_sec_NS} gives main characteristics of the laser and electron beams employed for producing LCS $\gamma$-ray beams at the NewSUBARU synchrotron.   
In Sec.~\ref{label_sec_MC_code} we give the theoretical modeling of the laser and electron beams and their Compton interaction, as well as the Monte Carlo simulation procedure and examples of flux and energy simulation results obtained with the \texttt{eliLaBr} code. 
In Sec.~\ref{label_comparison_with_exp} we compare our simulation results with NewSUBARU experimental LCS $\gamma$-ray beam spectra and study the effect of collimator offsets, small laser~--~electron beam misalignment and laser beam transverse distribution on the $\gamma$-ray flux, spectral distribution and maximum energy. 
In Sec.~\ref{label_sec_laser_pol}, we investigate the influence of laser polarization orientation on the LCS $\gamma$-ray beam properties, with focus on the spectral distribution and maximum energy edge preservation. Finally, conclusions are drawn in Sec.\ref{label_sec_conclusions}. 

\begin{figure*}[t]
\centering
\includegraphics[width=0.9\textwidth, angle=0]{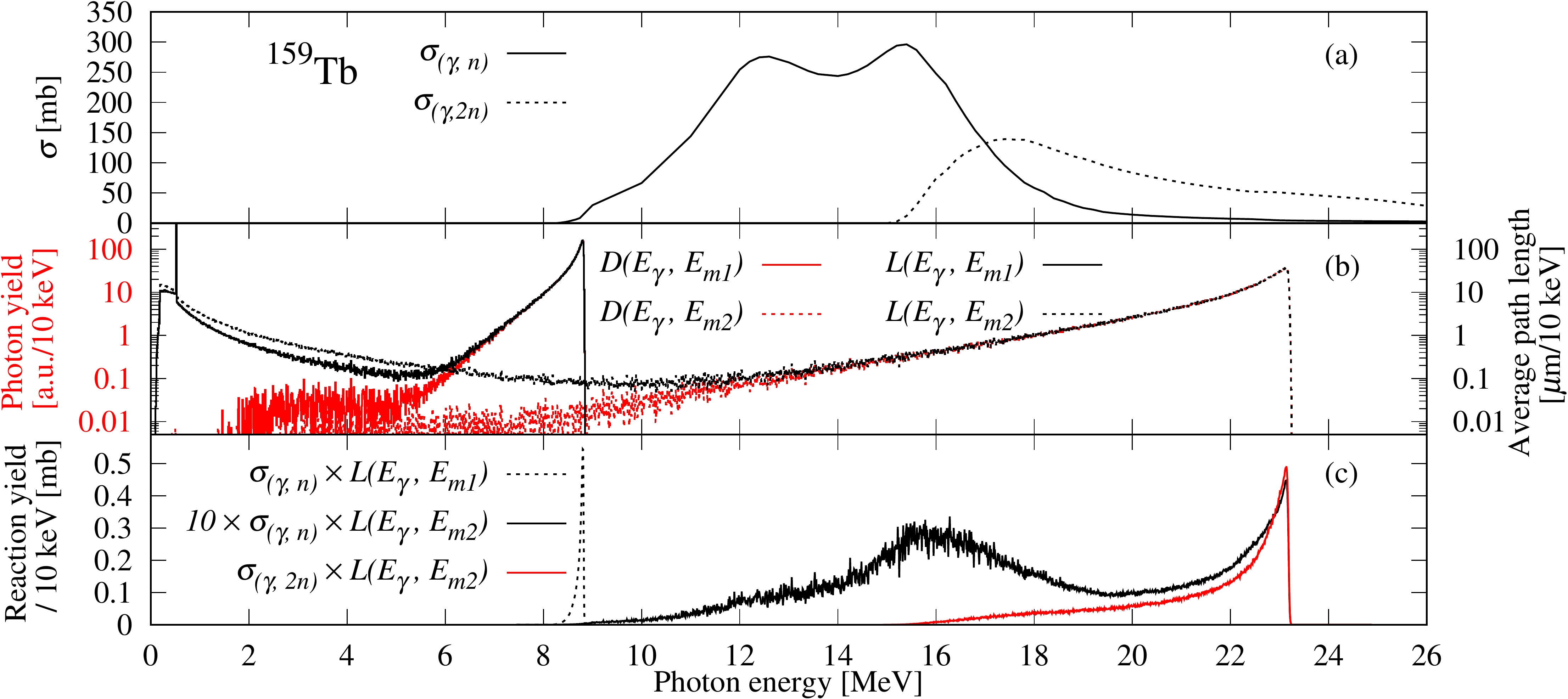}
\caption{(a) Evaluated data for the $^{159}$Tb$(\gamma,\,n)$ and $^{159}$Tb$(\gamma,\,2n)$ cross sections retrieved from the 2019 IAEA Photonuclear Data Library \cite{Kawano2020}. (b) Typical $D(E_\gamma,E_m)$ spectral distributions for NewSUBARU LCS $\gamma$-ray beams (red lines), where the $E_{m1}$~=~8.81~MeV maximum energy beam is produced with the 1048~nm wavelength INAZUMA laser and a 706.57~MeV electron beam and the $E_{m2}$~=~23.18~MeV one with the 524~nm Talon laser and 816.92~MeV electrons. The $L(E_\gamma,E_m)$ average path length distributions per unit energy in a 4~mm thick metallic Tb target are represented for both spectra in black lines. (c) The energy distribution of the $^{159}$Tb$(\gamma,\,n)$ and $^{159}$Tb$(\gamma,\,2n)$ cross sections folded by the $L(E_\gamma,E_m)$ distributions of the incident beams.   
}
\label{fig_FOLD}       
\end{figure*}

\section{Motivation - LCS $\gamma$-ray beam probing of photoneutron reactions in the GDR region}
\label{label_sec_motivation}

Recently, photonuclear reaction cross section measurements in the Giant Dipole Resonance (GDR) region have been performed at the $\gamma$-ray beamline of the NewSUBARU synchrotron radiation facility at SPring8, Japan. Photoneutron reactions $(\gamma,~xn)$ with $x$~=~1~to~4 have been investigated in the energy range between neutron emission threshold and $\sim$~42~MeV for nuclei in the wide mass range A~=~9~to~209~\cite{Gheorghe2017,Kawano2020}. Photofission experiments have been recently performed on $^{232}$Th and $^{238}$U~\cite{Filipescu_ND2022}, providing photoneutron and photofission cross sections, as well as prompt fission neutrons average energies and multiplicities.   
Neutron multiplicity sorting requirements have been met by developing a flat-efficiency neutron detection system along with associated sorting methods for both low~\cite{utsunomiyaNimDNM} and high-reaction rate conditions \cite{IGheorghe_2021_MF}. 

\subsection{LCS $\gamma$-ray beam diagnostics at NewSUBARU}

Key incident beam diagnostics requirements for absolute cross section measurements are met at the NewSUBARU facility, as following:

\emph{Absolute energy of the incident $\gamma$-ray beam.} The energy of the NewSUBARU electron beam has been calibrated by means of low-energy LCS beams produced using CO$_2$ laser photons \cite{IEEE_Utsunomiya14,Shima14}, with a relative uncertainty in the order of $10^{-5}$, typical for LCS electron beam energy measurements \cite{Klein2002,Sun2009_STAB,Chouffani2006}.  

\emph{Absolute flux of the incident $\gamma$-ray beam.} The pile-up, or Poisson fitting method~\cite{Toyokawa2000,kondo11,utsunomiya_nimMP_2018} is used for precise determination of incident photon fluxes for LCS $\gamma$-ray beams produced at NewSUBARU, as a function of time and with low uncertainty levels between 1$\%$ and 3$\%$. 

\emph{Spectral distribution of the incident $\gamma$-ray beam.} Quasi-monochromatic spectra are obtained by selecting through collimation the high-energy, backscattered component of the Compton spectrum. A large volume 3.5" diameter~$\times$~4.0" length LaBr$_3$:Ce detector is used for monitoring the incident $\gamma$-ray beam spectra at NewSUBARU. The procedure for the LCS $\gamma$-ray beam incident spectra diagnostics consists in the following steps: 
\begin{enumerate}
\item For each incident energy, LCS $\gamma$-ray beam energy spectra are recorded with the LaBr$_3$:Ce detector. The laser is operated in continuous wave mode at a reduced power, to avoid pile-up events. 
\item \label{step_2_diagnosis} The energy calibration of the LaBr$_3$:Ce detector response is performed using the maximum energy edge for each incident LCS $\gamma$-ray beam. The maximum LCS $\gamma$-ray energy is obtained for backscattered photons in head-on collision, and it is known from the energy of the laser photon beam and the energy of the calibrated electron beam.    
\item The incident spectral distributions are obtained by reproducing the calibrated experimental detector response by Monte Carlo simulations of LCS $\gamma$-ray beam production.
\end{enumerate}

We have previously shown in Refs.~\cite{IEEE_Utsunomiya14,filipescu_2014_sm} that second degree polynomials can well model the non-linearity effects in the 3.5"~$\times$~4.0" LaBr$_3$ detector response at medium energies up to $\sim$~15~MeV. We found that, when extending the energy range up to $\sim$~40~MeV, third degree polynomial functions fit best the experimental points in the energy calibration, as shown in Fig.~\ref{fig_labr_cal}. Gosta~{\it et al.}~\cite{gosta_2018} also used third degree polynomial functions in the 6~to~38~MeV energy range, and attributed the non-linearity effects to the photomultiplier tubes (PMT) coupled to large volume (3.5 inch $\times$ 8 inch) LaBr$_3$ crystals. 

The non-linear effects in the PMT response for tens of MeV $\gamma$ photons make the self-calibration by means of the maximum energy edge of the LCS $\gamma$-ray beams (step 2 above) a critical aspect in the diagnostics process. Thus, particular attention must be dedicated to the LCS $\gamma$-ray beam high energy edge preservation through good alignment conditions between the collimation system and the laser and electron beam axis. In the present work, we discuss the optimization of the alignment process in terms of the laser photon polarization. 

\begin{figure*}[t]
\centering
\includegraphics[width=0.85\textwidth, angle=0]{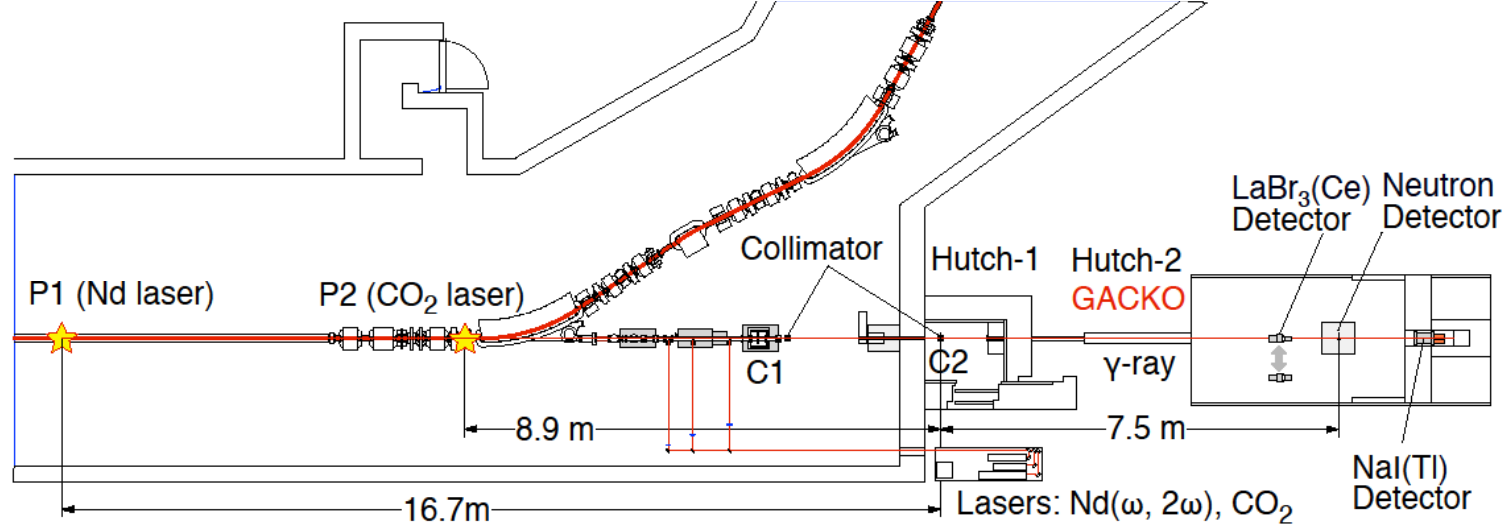}  
\caption{ (Color online) The Laser Compton scattering (LCS) $\gamma$-ray beamline BL01 of the NewSUBARU synchrotron radiation facility.    
}
\label{fig_NS_BL01}       
\end{figure*}

\subsection{Energy convolution of excitation functions}

Although the spectral density concentration in the maximum energy region of the spectra makes LCS $\gamma$-ray beams ideal tools for investigating photonuclear excitation functions, the measured cross section is in fact the convolution, or folding, between the true excitation function and the energy distribution of the incident beam, as shown for a test case in Fig.~\ref{fig_FOLD}. Here we consider the probing of the $^{159}$Tb $(\gamma,\,n)$ and $(\gamma,\,2n)$ cross sections shown in Fig.~\ref{fig_FOLD}(a) with a LCS $\gamma$-ray beam of $E_{m1}$~=~8.81~MeV maximum energy (full lines in Fig.~\ref{fig_FOLD}(b)), typical for low energy photoneutron cross section measurements in the vicinity of the neutron emission threshold $S_n$, and a $E_{m2}$~=~23.18~MeV maximum energy beam (dotted lines in Fig.~\ref{fig_FOLD}(b)) typical for neutron multiplicity sorting experiments above the two neutron separation energy $S_{2n}$. 

Conditioned that the energy spectra of the incident photon beams are not altered by the electromagnetic interaction of the incident beam with the target material, the measured yields are expressed as the folding between the true $\sigma_{\gamma,kn}(E_\gamma)$ cross sections and the $D(E_\gamma,E_{mi})$ energy distributions shown by red lines in Fig.~\ref{fig_FOLD}(b):
\begin{equation}\label{EQ_sigma_thin_t}
\sigma_{\gamma,kn}^{\mathrm{fold}}(E_{mi}) = \int D(E_\gamma,E_{mi})\cdot\sigma_{\gamma,kn}(E_\gamma)dE_\gamma,
\end{equation}
where we consider a normalized energy distribution with $\int_{S_{kn}}^{E_{mi}} D(E_\gamma,E_{mi}) dE_\gamma=1$. 

The above treatment has been successfully applied for photoneutron cross section measurements in the vicinity of the $S_n$, where the electromagnetic interaction of the incident beam generates secondary photons with energies below photonuclear reaction thresholds~\cite{TRenstrom_unfolding}. However, the spectral condition no longer holds with the increase in the maximum energy of the incident LCS $\gamma$-ray beam. Thus, one has to model the secondary photon production in the target and assess their contribution to the measured reaction yield. For this, we have found a useful tool in the $L(E_\gamma,E_m)$ distribution defined as the average path length per unit energy traveled through the target by a $E_\gamma$ photon in a LCS $\gamma$-ray beam of $E_m$ maximum energy. Fig.~\ref{fig_FOLD}(b) shows in black lines the average path length distributions for the two test LCS $\gamma$-ray beams, where, for the $E_{m2}$~=~23.18~MeV maximum energy beam, we notice a significant component of secondary photons with energies above the $S_n$. 

Using the $L(E_\gamma,E_m)$ distribution, we express the measured cross section $\sigma_{\gamma,kn}^{fold}$ as: 
\begin{equation} \label{EQ_CS_mono_LS}
\sigma_{\gamma, \,kn}^{\mathrm{fold}}(E_{m})=\cfrac{1}{\xi} \int_0^{E_{m}} L(E_\gamma,E_m) \sigma_{\gamma,kn}(E_\gamma)\,dE,   
\end{equation} 
where $\xi=(1-e^{-\mu L})/\mu$ is a self attenuation correction factor determined by the linear attenuation coefficient $\mu$ and target thickness $L$, as described in~Ref.~\cite{IGheorghe_2021_MF}. 
By approximating the continuous $L(E_\gamma,E_m)$ and $\sigma_{\gamma,\, kn}$ functions in Eq.~\ref{EQ_CS_mono_LS} with discrete sequences, with a typical energy grid of $dE$~=~10~keV, we obtain:
\begin{equation} \label{EQ_matrix_form_fold}
\sigma_{\gamma, \,kn}^{\mathrm{fold}}(E_m)= \cfrac{1}{\xi} \, \sum_{i=0}^{E_m/dE} L(i,E_m)\sigma_{\gamma, \,kn}(i)
\end{equation}
where the index $i$ of $L(i,E_m)$, $\sigma_{\gamma, \,kn}(i)$ and $\sigma_{\gamma, \,kn}^{\mathrm{fold}}(i,E_m)$ stands for a $E_\gamma$~=~$i\,\cdot\,dE$ photon energy. 

Fig.~\ref{fig_FOLD}(c) shows the folding between the $(\gamma,\,n)$ and $(\gamma,\,2n)$ cross sections of Fig.~\ref{fig_FOLD}(a) with the $L(E_\gamma,E_m)$ distributions of Fig.~\ref{fig_FOLD}(b). The low energy beam interrogates the cross section on the increasing slope of the GDR, thus the photoneutron yield is mostly generated in the vicinity of the 8.81~MeV maximum energy. The 23.18~MeV maximum energy beam interrogates the cross section on the decreasing GDR slope. As the high energy GDR peak region folds with the low energy tail of the LCS beam, large fractions of the measured photoneutron reaction yields are generated by the tail of the energy distribution, especially for the $(\gamma,~n)$ channel. The experimental excitation function is obtained by unfolding the measured cross section, following the iterative procedure described in Ref.\cite{TRenstrom_unfolding}. It is seen that the accuracy of the unfolded results is highly dependent on the good characterization of the LCS $\gamma$-ray spectra.

\begin{figure}[t]
\centering
\includegraphics[width=0.45\textwidth, angle=0]{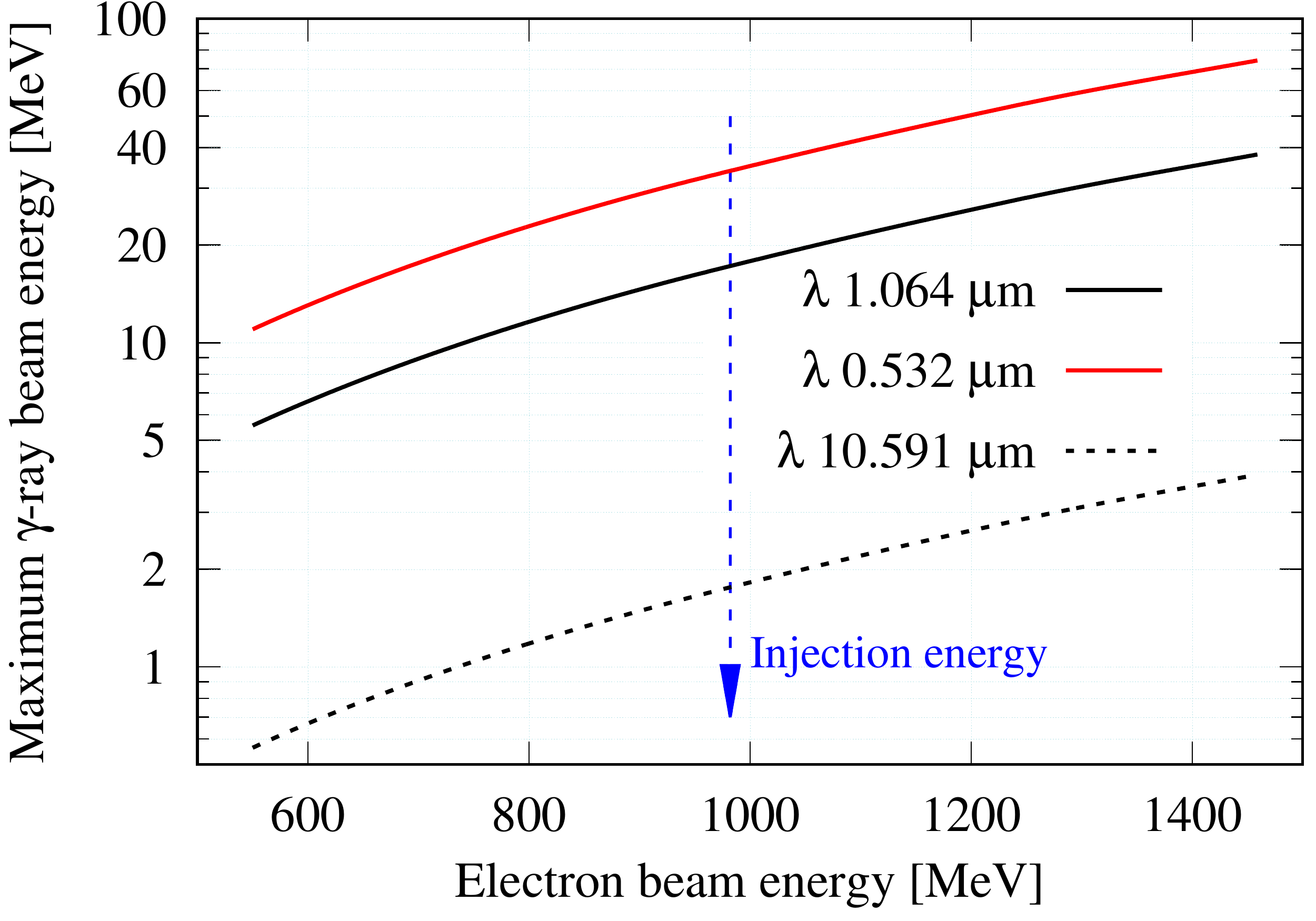} 
\caption{Maximum energy of LCS $\gamma$-ray beams produced at NewSUBARU with the INAZUMA laser of $\lambda$~=~1.064~$\mu$m, Talon laser of $\lambda$~=~0.532~$\mu$m and CO$_2$ laser of $\lambda$~=~10.59~$\mu$m.   
}
\label{fig_max_g_en}       
\end{figure}

\section{NewSUBARU LCS $\gamma$-ray beamline}
\label{label_sec_NS}

The experimental setup for the recent photoneutron experimental campaign at the beamline BL01 of the NewSUBARU synchrotron radiation facility is schematically shown in Fig.~\ref{fig_NS_BL01}. Laser beams are sent head-on against electron beams circulating along a 20~m long straight section of the storage ring. The backscattered $\gamma$-ray beam is passed through a double collimation system and sent downstream in the experimental Hutch-2 GACKO (Gamma Collaboration Hutch of Konan University), where the target, neutron detection system and flux and energy profile monitoring systems are placed. Here we give information on the NewSUBARU LCS $\gamma$-ray source relevant for our present study, and guide the reader to Refs.~\cite{aoki2004,miyamoto2007,amano09,Horikawa2010,IEEE_Utsunomiya14} for further details.  

\subsection{Maximum LCS $\gamma$-ray energy}

For Compton scattering of laser photons against relativistic electrons, the energy corresponding to backscattered photons in head-on collisions is given by:
\begin{equation}
E_\gamma=\cfrac{4\gamma^2E_p}{1+(\gamma\theta)^2+4\gamma E_p/(mc^2)},
\end{equation}
where $E_p$ is the laser photon energy, $mc^2$ is the electron rest mass, $\theta$ is the scattering angle of the $\gamma$-ray photon relative to the electron incident direction, and $\gamma$ is the Lorentz factor for the electron. 
Thus, the maximum energy of the LCS $\gamma$-ray beam, for $\theta$~=~0, is directly determined by the energy of the electron beam and of the laser photon. 

At NewSUBARU, the energy of the electron beams can be continuously varied between 0.5 and 1.5~GeV in either deceleration or acceleration mode starting from the 982 MeV injection energy. $\gamma$-ray beams with maximum energies between 0.5~MeV and 76~MeV can be produced by using lasers of different wavelengths: a 10.591~$\mu$m wavelength CO$_2$ laser and two diode-pumped solid state lasers, the 1.064~$\mu$m Nd:YVO$_4$ INAZUMA laser and the 0.532~$\mu$m Talon laser. Figure~\ref{fig_max_g_en} shows the maximum $\gamma$-ray energy corresponding for each of the three lasers for the entire electron beam energy range. 

\begin{figure}[t]
\centering
\includegraphics[width=0.45\textwidth, angle=0]{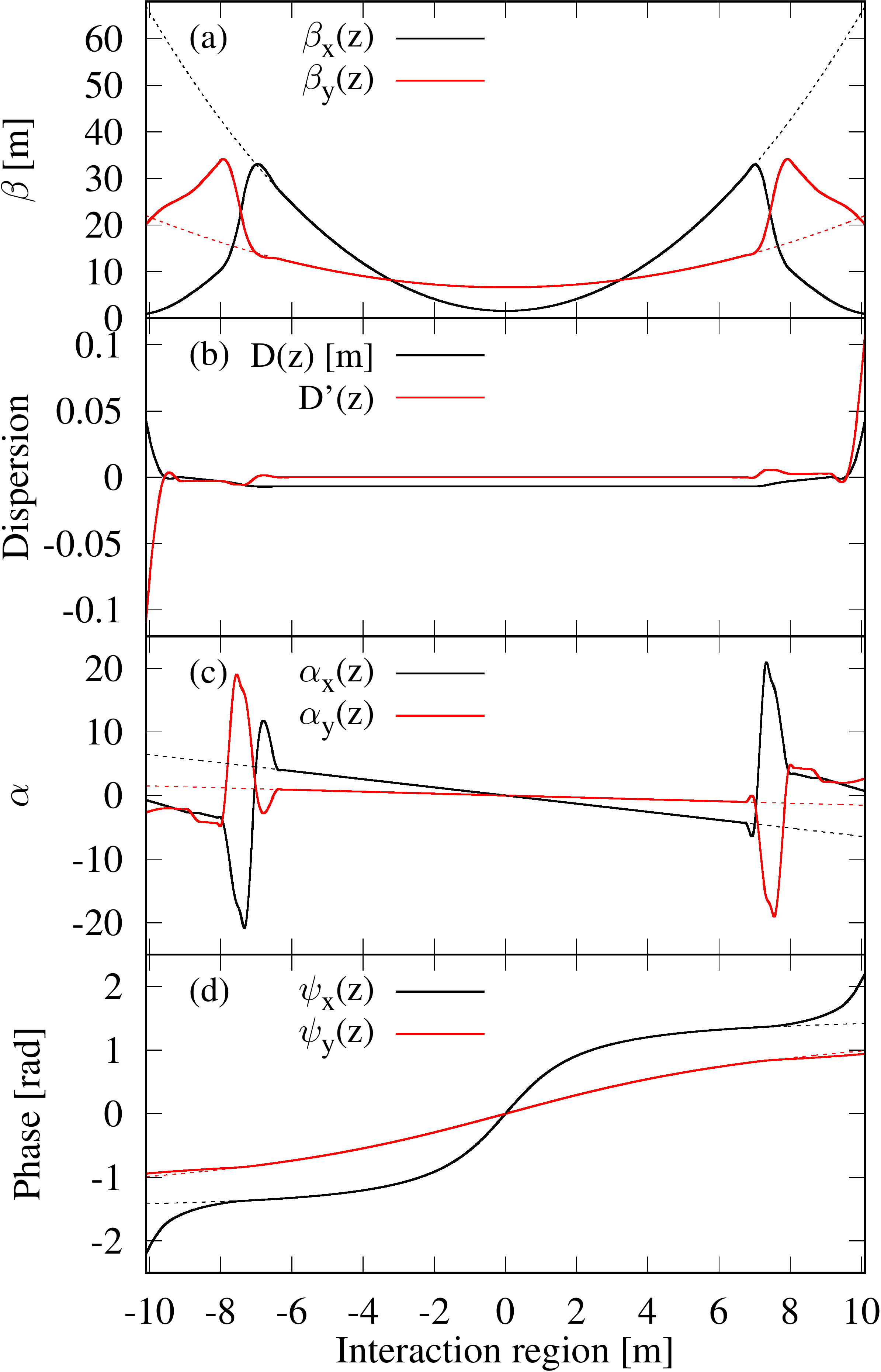} 
\caption{Electron beam parameters for the long straight section of the NewSUBARU ring: Twiss parameters (a) $\beta$ and  (c) $\alpha$, (b) horizontal axis dispersion functions $D$ and $D$, (d) phase advance function $\psi$.   
}
\label{fig_lattice_par}       
\end{figure}

\subsection{Collimation}

Quasi-monochromatic $\gamma$-ray beams corresponding to the maximum energy component of the Compton spectrum are obtained by angular selection of the backscattered photon spectra through the double collimation system shown in Fig.~\ref{fig_NS_BL01}. The first collimator (C1) is placed inside the electron beam vault at 15.47~m away from the center of the electron beamline. The second collimator (C2) is placed in the experimental Hutch-1 at 18.47~m from the center of the electron beamline. The beam energy spectrum monitor and the target are placed at $\sim$~6.5~m, respectively $\sim$7.5~m downstream of the C2 collimator. Each collimator is made of 10-cm-thick Pb blocks. In our measurements, we have used apertures of 6~mm and 3~mm for C1 and 2~mm and 1~mm for C2. The collimators are mounted on $x$-$y$-$\theta$ stages driven by stepping motors which allow finetunings of their alignment on the horizontal ($x$), vertical ($y$) and rotational axes ($\theta$).  

\subsection{Electron and laser beam time profile}

At NewSUBARU, the electron beam bunches have a 500~MHz frequency (2~ns interval) and 60~ps width. 
The laser is operated in Q-switch mode at frequencies between 1 and 25~kHz, producing photon pulses 40 to 60~ns wide. 
Because the electron and laser beams are unsynchronized, they can be modeled as continuous beams with constant densities of laser photons and electrons along the beamline.  

\subsection{Electron beam spatial profile}

The spatial profile of the electron beam is characterized by Gaussian distributions in the horizontal and vertical axes. The standard deviations $\sigma_x(z)$ and $\sigma_y(z)$ at the $z$ position along the electron beam axis are described by
\begin{equation}\label{label_eq_sigma_fd_beta}
\sigma_x(z)=\sqrt{\beta_x(z) \varepsilon_x+(\Delta E/E_0)^2D_x^2(z)} \, , \, \sigma_y(z)=\sqrt{\beta_y(z)\varepsilon_y},
\end{equation}
where $\varepsilon_{x,y}$ are the horizontal and vertical emittances, $\beta_{x,y}$ the betatron functions, $D_x$ the horizontal dispersion function and $\Delta E/E_0$ the energy spread of the electron beam. For the injection energy of 982.43~MeV, the nominal value for the rms energy spread is 0.04$\%$.  
The natural emittance at injection energy is 38~nm-rad with an XY coupling less than 1\%. An RF shaker composed of four strip-line electrodes with RF frequency equal to betatron frequency is used at NewSUBARU, to intentionally enlarge the vertical beam size and improve the Touschek lifetime. As a result, the measured vertical beam size is a few times larger than that without RF shaker \cite{utsunomiya_2015_npn}.

\begin{figure}[t]
\centering
\includegraphics[width=0.45\textwidth, angle=0]{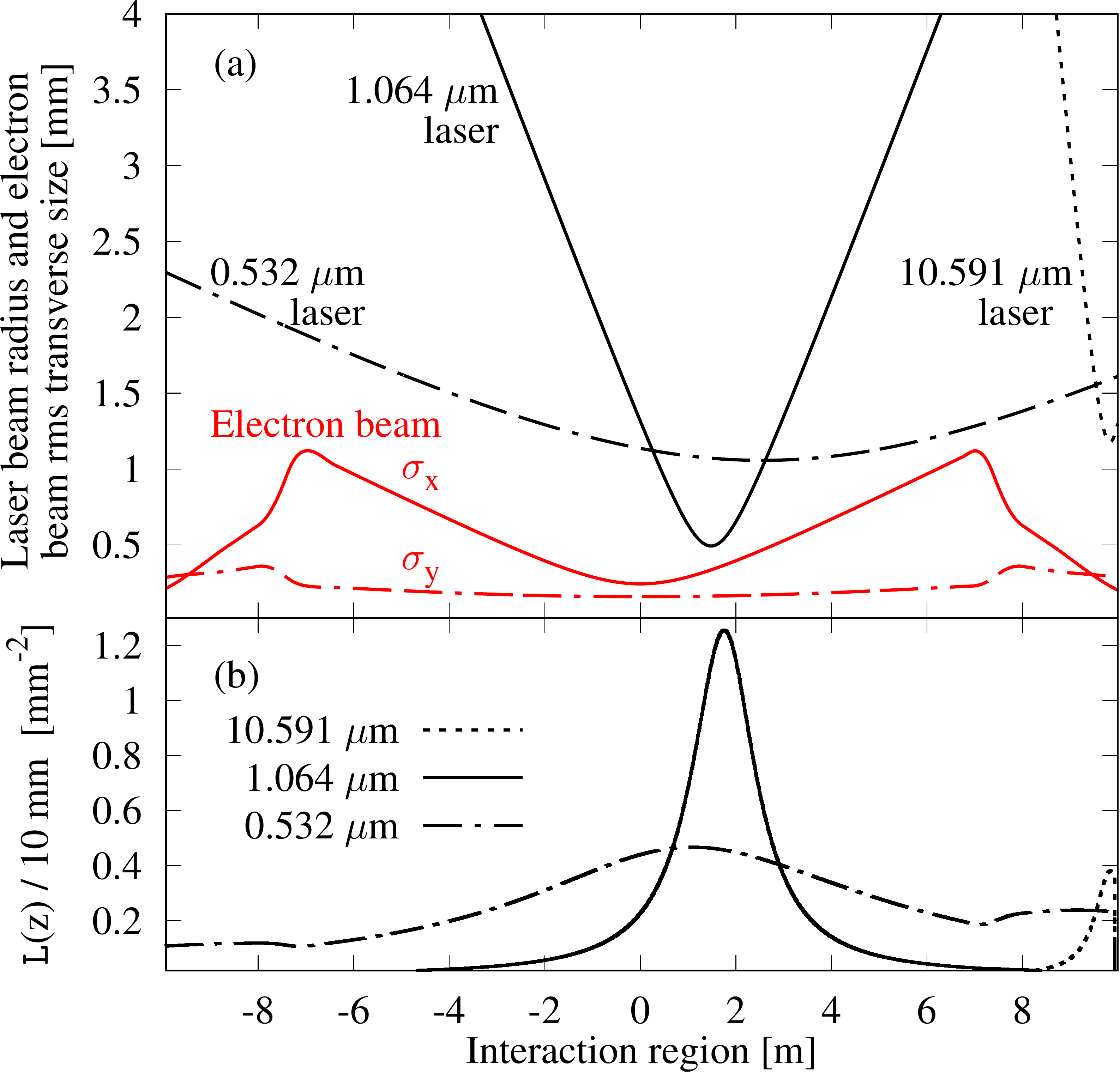} 
\caption{(a) Electron beam rms transverse size on horizontal ($\sigma_x$ full red line) and vertical ($\sigma_y$ dashed dotted red line) axes given by Eq.~\ref{label_eq_sigma_fd_beta} using MAD-X calculated beam parameters shown in Fig.~\ref{fig_lattice_par} and considering horizontal $\varepsilon_x$~=~38~nm-rad and vertical $\varepsilon_y$~=~3.8~nm-rad emittance. Gaussian optics calculations of the transverse beam radii for the INAZUMA laser of $\lambda$~=~1.064~$\mu$m (full black line), Talon laser of $\lambda$~=~0.532~$\mu$m (dashed dotted line) and CO$_2$ laser of $\lambda$~=~10.591~$\mu$m (dotted black line). (b) Overlap function between lasers and electron beams, as defined in Eq.~\ref{label_eq_luminosity_approximate_particular}.   
}  
\label{fig_beam_rad}       
\end{figure}

Fig.~\ref{fig_lattice_par} shows the Twiss parameters $\alpha$ and $\beta$, $\psi$ betatron phase advance function and $D$, $D'$ dispersion functions along the long straight section of the NewSUBARU ring calculated with the MAD-X code~\cite{mad_x} (solid lines) and compared with their corresponding values for a free electron drift section (dashed line). Fig.~\ref{fig_beam_rad}(a) shows the transverse size of the electron beam on horizontal (full red line) and vertical (dashed dotted red line) axes in the interaction region calculated using the MAD-X results given in Fig.~\ref{fig_lattice_par}. 
We note that the electron beam focus lies in the center of the beamline, referred to as the zero value on the horizontal axis of Fig.~\ref{fig_beam_rad}(a). 


\subsection{Laser beam spatial profile}

The transverse intensity profile of the laser beams is characterized by Gaussian functions in the horizontal and vertical directions with a symmetrical standard deviation $\sigma_p$, which represents half of the laser beam radius, accordingly to the $D4\sigma$ definition of the laser diameter. For Gaussian beams, the transverse width $\sigma_p$ at a $z_p$ position along the laser propagation axis is expressed as:
\begin{equation}\label{label_eq_sigma_p_fd_Zr}
\sigma_p(z_p)={D_{p0}}/{4} \cdot \sqrt{1+\left({z_p}/{Z_R} \right)^2},
\end{equation}
where $D_{p0}$ is the laser diameter in the laser waist. $Z_R$ is the Rayleigh length defined by the laser waist diameter, laser wavelength $\lambda$ and laser quality factor $M^2$, as following:
\begin{equation}
Z_R=\cfrac{\pi D_{p0}^2}{4M^2\lambda}. 
\end{equation}

Fig.~\ref{fig_beam_rad}(a) shows the CO$_2$, INAZUMA and Talon laser beams radii along the interaction region, which were obtained with the thin lens equation for Gaussian beams by considering the geometrical distances and the specifications for the laser (wavelength and diameter at laser output) and the focusing lens focal length. We note the increase in the divergence of the laser beam with the laser wavelength. The CO$_2$ (10.591 $\mu$m) and the INAZUMA (1.064 $\mu$m) lasers are well localized at their corresponding focal points, with the CO$_2$ focus placed in the region of the quadrupole magnet. The low wavelength Talon laser (0.532 $\mu$m) shows a much smoother diameter dependence along the electron beamline. The main parameters for the three lasers, such as the wavelength, diameter at beam waist, Rayleigh length and the displacement of the laser focus relative to the electron beam focus along the beam axis are given in Table~\ref{tab_lasers}.  

\begin{table}[t]
\begin{center}
\caption {Parameters for the lasers used in the present study: wavelength ($\lambda$), diameter ($D_{p0}$) at focus, Rayleigh length ($Z_R$) and displacement between the laser and electron beam focus points ($\Delta z$).} 
\label{tab_lasers}
  \begin{tabular}{  l | r   r   r   r }
  \hline \hline
    Laser    &  $\lambda$ [$\mu$m]       & $D_{p0}$ [mm] & $Z_R$ [cm] & $\Delta z$ [m]     \\ \hline          
    CO$_2$   &  10.591                   & 1.7             & 20.        & 9.8              \\ 
    INAZUMA  &  1.064                    & 1.0             & 57.        & 1.8              \\     
    Talon    &  0.532                    & 2.1             & 6100.      & 2.8              \\ \hline \hline       
  \end{tabular}
\end{center}
\end{table}

\section{Monte Carlo modeling of LCS $\gamma$-ray beam sources}
\label{label_sec_MC_code}

\subsection{Modeling of laser and electron beams}
\label{label_subsec_modeling_of_laser_elecron_beam}

Continuous laser and electron beams are considered. The electron beam propagates along the $z$ direction in the $(x,y,z)$ coordinate system. $(x_p,y_p,z_p)$ is the coordinate system for the laser beam propagating in the negative $z_p$ direction. For both electron and laser beams, the center of the beam focal waist is at $(0,0,0)$ in the respective coordinate system. 

Considering a Gaussian electron beam transverse distribution, the electron density $dN_e(x,y,z)$ is given by \cite{Horikawa2010}:
\begin{equation}\label{label_eq_dNe}
dN_e(x,y,z) = \cfrac{I}{ev} \cdot dn_e(x,y,z),
\end{equation}
where $I$ is the electron beam current, $e$ and $v$ are the electron charge and velocity, and 
\begin{equation}\label{label_eq_dNe}
dn_e(x,y,z) = \cfrac{1}{2\pi\sigma_x(z)\sigma_y(z)}\exp\left[-\cfrac{1}{2}\left(\cfrac{x}{\sigma_x(z)}\right)^2-\cfrac{1}{2}\left(\cfrac{y}{\sigma_y(z)}\right)^2\right].
\end{equation}

Assuming the laser to be a Gaussian beam, the laser photon density $dN_p(x_p,y_p,z_p)$ is given by:
\begin{equation}\label{label_eq_dNp}
dN_p(x_p,y_p,z_p)=\cfrac{P}{E_p c} \cdot dn_p(x_p,y_p,z_p),
\end{equation}
where $P$ is the laser power, $E_p$ is the energy of the laser photon, $c$ is the speed of light, and
\begin{equation}\label{label_eq_dNp_overlap}
dn_p(x_p,y_p,z_p)=\cfrac{1}{2\pi\sigma_p^2}\exp\left(-\cfrac{1}{2}\cfrac{x_p^2+y_p^2}{\sigma_p^2} \right).
\end{equation}

\subsection{Flux of Compton scattered $\gamma$-ray beam}
\label{label_subsec_luminosity}

We express the number of Compton scattered $\gamma$-rays per unit time by:
\begin{equation}\label{label_eq_dNg_dt}
\cfrac{dN_\gamma}{dt} = \mathcal{L}\cdot\sigma_{tot},
\end{equation}
where $\sigma_{tot}$ is the angle integrated cross section for the Compton scattering of laser photons on relativistic electrons for head-on collisions (see Eq.~(15) of Ref.~\cite{Sun2011_STAB}). $\mathcal{L}$ is the luminosity, expressing the number of scattering events produced per unit scattering cross section and per unit time and defined as \cite{WernerMuratori2006}:
\begin{equation}\label{label_eq_luminosity_general}
\mathcal{L} = c(1+\beta)\cdot \int_x\int_y\int_z dN_e dN^*_p dx dy dz,
\end{equation}
where $dN^*_p$ is the laser density function of Eq.~\ref{label_eq_dNp} expressed in the $(x,y,z)$ coordinate system in which the electron beam propagates along the $z$ axis and $c(1+\beta)$ is the kinematic factor for head-on collisions between laser photons and relativistic electrons with velocity $v$~=~$\beta\cdot c$. 

At NewSUBARU, the laser beam is aligned to the electron beam axis for a head-on collision geometry. There is generally a non-zero $\Delta z$ displacement between the electron beam focus point found in the middle of the long straight section of the accelerator and the laser focal position determined by the optics system and laser parameters (wavelength, beam spot, quality factor M$^2$). Besides the $\Delta z$ displacement, we also consider small transverse off-sets and rotations between the laser and electron beam axis generated by non-ideal alignment settings, which can have significant impact on the spectral distribution and incident flux of the LCS $\gamma$-ray beams. Thus, considering $(x_p,y_p,z_p)$ a coordinate in the laser system, we perform the following transformations:
\begin{itemize}
\item a rotation of $\theta$ and $\varphi$ spherical angles of the laser axis relative the electron axis: 
\begin{equation}\label{eq_rotation}
(x'_p,y'_p,z'_p)=\mathcal{R}_{\theta_y, \varphi_z} (x_p,y_p,z_p);
\end{equation}
\item followed by a spatial displacement $\Delta x$, $\Delta y$, $\Delta z$ of the laser through the translation operator:
\begin{equation}\label{eq_translation}
(x,y,z) = \mathcal{T}_{\Delta x, \Delta y, \Delta z}(x'_p,y'_p,z'_p)
\end{equation}
\end{itemize}
Now we can define :
\begin{align}\label{label_eq_dNp_star}
\begin{split}
dN^*_p(x,y,z) & = dN_p(x_p,y_p,z_p), \\
dn^*_p(x,y,z) & = dn_p(x_p,y_p,z_p). 
\end{split}
\end{align}

Under the condition of small angle rotations, the transverse laser intensity distribution in point $z$ along the electron beam axis can still be approximated to a Gaussian distribution, with standard deviation $\sigma^*_p(z)$~$\approx$~$\sigma_p(z-\Delta z)$. 

To compute Eq.~\ref{label_eq_luminosity_general}, we introduce a laser~--~electron overlap function given by the following integral over the transverse coordinates in each $z$ position along the electron beam axis: 
\begin{equation}\label{label_eq_luminosity_approximate_general}
\texttt{L}(z)=\int_x\int_y dn_e(x,y,z) dn_p(x_p,y_p,z_p) dx dy,
\end{equation}
given explicitly as:
\begin{equation}\label{label_eq_luminosity_approximate_general_expl}
\texttt{L}(z)= \cfrac{ \exp\left(-\cfrac{x^2_{p0}(z)}{2\big(\sigma^2_{x}(z)+\sigma^{*2}_p(z)\big)} - \cfrac{y^2_{p0}(z)}{2\big(\sigma^2_{y}(z)+\sigma^{*2}_p(z)\big)}\right) }{2\pi \sqrt{\sigma^2_{x}(z)+\sigma^{*2}_p(z)}\cdot\sqrt{\sigma^2_{y}(z)+\sigma^{*2}_p(z)} },
\end{equation}
where $x_{p0}(z)$ and $y_{p0}(z)$ are the transverse coordinates of the laser axis relative to the electron axis at $z$, obtained through the small angle rotation $\mathcal{R}_{\theta_y, \varphi_z}$ followed by $\mathcal{T}_{\Delta x, \Delta y, \Delta z}$ laser translation. For an off-axis laser beam with no rotations relative to the electron axis, the $x_{p0}$ and $y_{p0}$ transverse displacements are constant along the $z$ axis, with $x_{p0}(z)$~=~$\Delta x$ and $y_{p0}(z)$~=~$\Delta y$. 

In case of perfectly aligned electron and laser beam propagating axes, with no transverse displacements and no rotations, Eq.~\ref{label_eq_luminosity_approximate_general_expl} is simplified as follows:
\begin{equation}\label{label_eq_luminosity_approximate_particular}
\texttt{L}(z)=\cfrac{ 1 }{2\pi \sqrt{\sigma^2_{x}(z)+\sigma^{*2}_p(z)}\cdot\sqrt{\sigma^2_{y}(z)+\sigma^{*2}_p(z)} }.
\end{equation}
Fig.~\ref{fig_beam_rad}(b) shows the overlap between the electron beam and the three laser beams of 10.591~$\mu$m, 1.064~$\mu$m and 0.532 $\mu$m wavelength, computed with Eq.~\ref{label_eq_luminosity_approximate_particular}. We note that, for the CO$_2$ and INAZUMA lasers, the effective interaction region is well localized at the laser focal point, while for the Talon laser the interaction takes place along the entire beamline. Finally, we further integrate Eq.~\ref{label_eq_luminosity_approximate_particular} on $z$ axis in order to obtain the luminosity:
\begin{equation}\label{label_eq_luminosity_aligned}
\mathcal{L} = c(1+\beta\cdot \cos\theta)\cdot\cfrac{I}{ev}\cfrac{P}{E_p c} \int_z \texttt{L}(z) dz,
\end{equation}
where we added the $\cos \theta$ multiplication into the kinematic factor to take into account the angle between the laser and the electron beams. The analytical estimation given by Ref.~\cite{Horikawa2010} for the NewSUBARU LCS $\gamma$-ray beam luminosity in the INAZUMA laser and 982.43~MeV electron energy configuration is reproduced by the present Monte Carlo modeling, considering transverse emittance values $\varepsilon_{x}$~=~78~nm-rad, $\varepsilon_{y}$~=~7.8~nm-rad and standard INAZUMA laser parameters given in Table~\ref{tab_lasers}. Our calculations estimate for this configuration a total normalized luminosity of 5.26~$\cdot$~10$^4$ $\gamma$-rays / (barn W mA s), which after being transported in \textsc{Geant4} by taking into account the vacuum beam pipe, the $\gamma$ absorber within the beam pipe, the transmission through laser mirror and vacuum window, reproduces the normalized photon flux of 2.1~$\cdot$~10$^4$ $\gamma$-rays / (W mA s) given in Ref.~\cite{Horikawa2010}, registered on a monitor screen placed in the experimental hutch. 

\begin{figure*}[t]
\centering
\includegraphics[width=0.85\textwidth, angle=0]{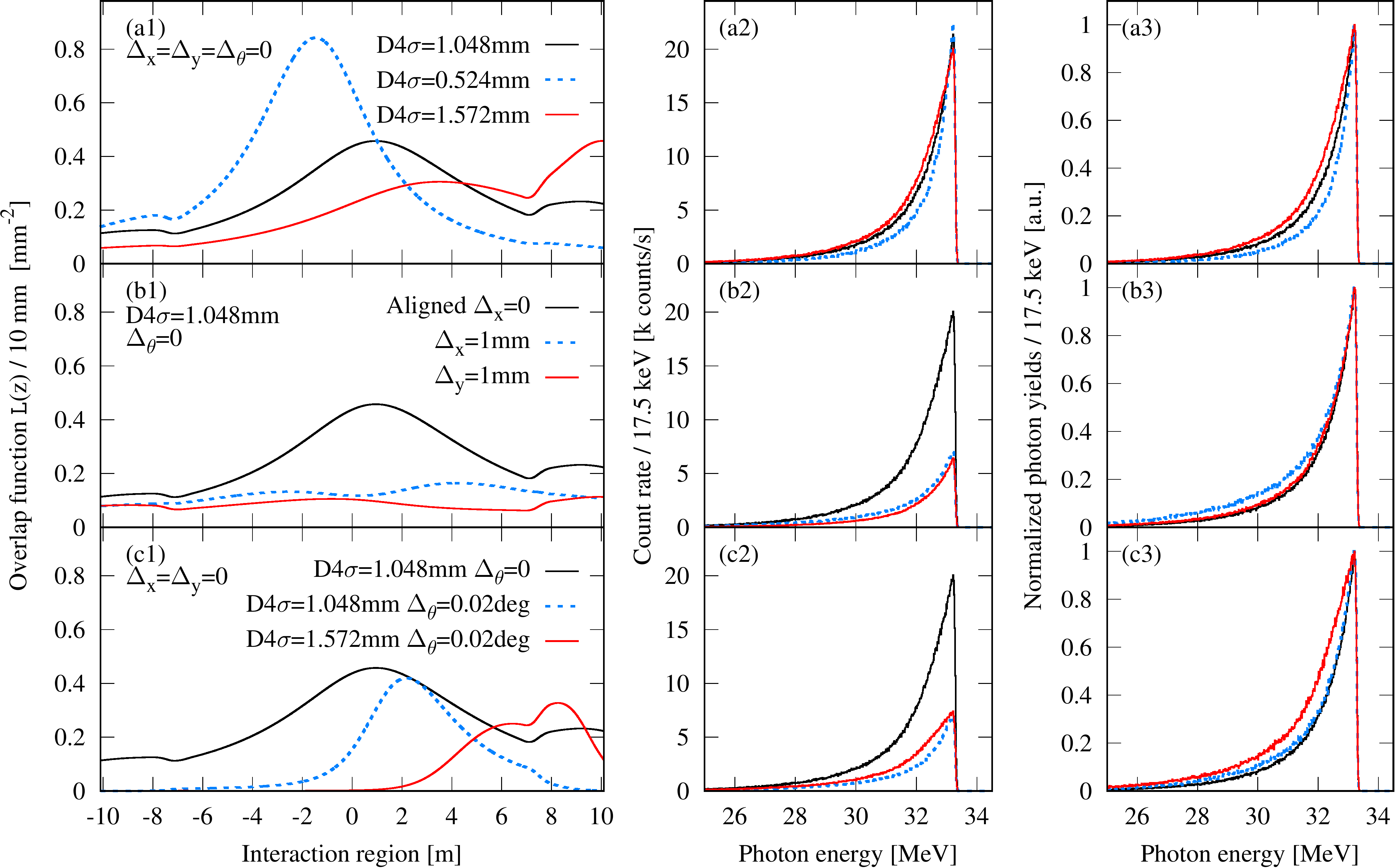} 
\caption{Simulations of LCS $\gamma$-ray beam spectra and incident flux for $E_e$~=~982~MeV and different configurations of the Talon laser: (a1--3) study for different laser beam diameter values D4$\sigma$ at laser output, where the nominal D4$\sigma$~=~1.048~mm; (b1--3) study for vertical and horizontal laser off-sets relative to the electron axis; (c1--3) study different D4$\sigma$ values and small $\theta$ rotation of laser relative to the electron axis. Good electron beam emittance values have been considered: $\varepsilon_x$~=~37~nm~rad, $\varepsilon_y$~=~3.7~nm~rad. Left column: overlap between laser and electron beam along the interaction region, as defined by Eq.~\ref{label_eq_luminosity_approximate_general_expl}. Middle column: count rate for collimated incident photon spectra. Right column: normalized incident photon spectra, for comparison of bandwidth.    
}
\label{fig_sim_spot}       
\end{figure*}

\subsection{Sampling procedure}
\label{label_subsec_sampling}

\subsubsection{Interaction point $z$ sampling }\label{label_sub_sub_sec_sample_z}
The $z$ laser~--~electron interaction position along the electron beam axis is sampled by the \emph{acceptance~--~rejection} method, accordingly to the laser beam -- electron beam overlap function $\texttt{L}(z)$ given by Eq.~\ref{label_eq_luminosity_approximate_general_expl}.
\subsubsection{Interaction point transverse coordinates sampling}\label{label_sub_sub_sec_sample_xy}
The ($x$, $x'$, $y$, $y'$)$_{z~=~0}$ transverse electron phase-space coordinates are sampled in the $z$~=~0 electron beam waist following the procedures introduced in Refs.\cite{CAIN} and \cite{Sun2011_STAB} (Eqs.~56), where we assume a Gaussian phase-space distribution and we take into consideration that the focal point of the electron beam is found in a free-electron drift section of the electron ring. Here $x'$~=~$dx/dz$ and $y'$~=~$dy/dz$ are the angular divergences of the incident electron. The electron phase-space coordinates at the $z$ sampled position along the interaction region are afterwards computed using the $\mathbf{M}$ transfer matrix for particle trajectories \cite{HelmutWiedemann_PAP}:
\begin{equation} \label{EQ_sis_lin_eq_Fold}
\begin{pmatrix}
    u \\
    u' 
\end{pmatrix}_{z}
=
\mathbf{M}
\cdot
\begin{pmatrix}
    u \\
    u' \\
    \Delta p/p
\end{pmatrix}_{z=0}
\end{equation}
where $u$ and $u'$ stand for either the horizontal $x$, $x'$ or vertical $y$, $y'$ transverse coordinates. The transfer matrix $\mathbf{M}$ is completely defined by the $\alpha$ and $\beta$ Twiss parameters, $\psi$ betatron phase advance function and $D$, $D'$ dispersion functions:
\begin{equation}
\mathbf{M}
=
\begin{pmatrix}
    \cfrac {\sqrt{\beta}(\cos\psi+\alpha_0\sin\psi)}{\sqrt{\beta_0}}                                         &\!\!\!\sqrt{\beta \beta_0}\sin\psi                      &\!\!D \\
    \cfrac{(\alpha_0\!-\!\alpha)\cos\psi\!-\!(\!1\!+\!\alpha_0\alpha)\sin\psi }{\sqrt{\beta \beta_0}} &\!\!\!\cfrac { \sqrt{\beta_0}(\cos\psi\!-\!\alpha\sin\psi) }  {\sqrt{\beta}} &\!\!D' 
\end{pmatrix}
\end{equation}
where we note that the vertical dispersion $D_{y}(z)$~=~$D'_{y}(z)$~=~0 along the storage ring and $\alpha_0$~=~0 in the focus.

Based on the transverse laser intensity distribution at sampled $z$ interaction point position, we decide whether to accept the sampled ($x$, $x'$, $y$, $y'$)$_{z}$ coordinates, by applying the $acceptance$~-~$rejection$ method on the $dN^*_p$ intensity distribution function given by Eqs.~\ref{label_eq_dNp_star},\ref{label_eq_dNp}. If the transverse sampled coordinates are not accepted, the code repeats Sec.~\ref{label_sub_sub_sec_sample_xy}. 

If the sampled transverse coordinates are accepted, the code generates the electron and laser quadrivectors in the sampled interaction point. The energies of the electron and photon are sampled accordingly to Gaussian distributions of $\Delta E_e$ and $\Delta E_p$ rms distributions, respectively. The momentum of the laser photon at the $(x_p,y_p,z_p)$ transformed from the sampled $(x,y,z)$ is obtained from the gradient of the laser propagation phase, considering a Gaussian laser beam \cite{CAIN,Sun2011_STAB}. 

\subsubsection{Scattered $\gamma$-ray photon sampling}\label{label_sub_sub_sec_sample_g}

Following the procedure applied in various existing Monte Carlo simulation codes of laser Compton scattering sources, such as \textsc{cain}~\cite{CAIN}, \textsc{mccmpt}~\cite{Sun2011_STAB}, \textsc{cmcc}~\cite{Curatolo2017,Curatolo_PhD_thesis}, etc., the electron and laser momenta quadrivectors are rotated and Lorentz transformed to obtain a head-on collision along the $z$ axis in the electron rest frame coordinate system. 

In the standard treatment of Compton scattering of polarized photons on unpolarized relativistic electrons~\cite{Sun2011_STAB,Luo2011,Curatolo2017,Hajima2021}, the sampling procedure accounts for the correlation between the scattering direction and the direction of the laser photon polarization vector, while the polarization properties of the scattered photon are described analytically~\cite{petrillo2015}. In the present algorithm, the scattered photon polarization modeling is implemented in the Monte Carlo method. As described in~\cite{Filipescu_2022_POL}, the transformations applied to the laser momentum quadrivector are also applied to the laser polarization quadrivector, followed by a gauge invariance transformation and the extraction of the Stokes polarization vector~\cite{McMaster1,McMaster2}. In Ref.~\cite{Filipescu_2022_POL}, the scattered photon polarization spatial distributions obtained with the \texttt{eliLaBr} code are validated against the analytical results given in Refs.~\cite{Sun2011_STAB,petrillo2015,zhijun_chi2020} and the fundamental harmonic results from Ref.~\cite{zhijun_chi2022}. 

Thus, the momentum quadrivector of the scattered $\gamma$-ray photon is generated accordingly to the well-known differential Klein-Nishina cross section for head-on collisions in the electron rest frame coordinate system. Also, the Stokes polarization vector is transformed accordingly to the Landau prescription given in~\cite{Landau}. Finally, the event is Lorentz transformed to the laboratory frame and rotated back to the initial orientation.   

\begin{figure}[t]
\centering
\includegraphics[width=0.45\textwidth, angle=0]{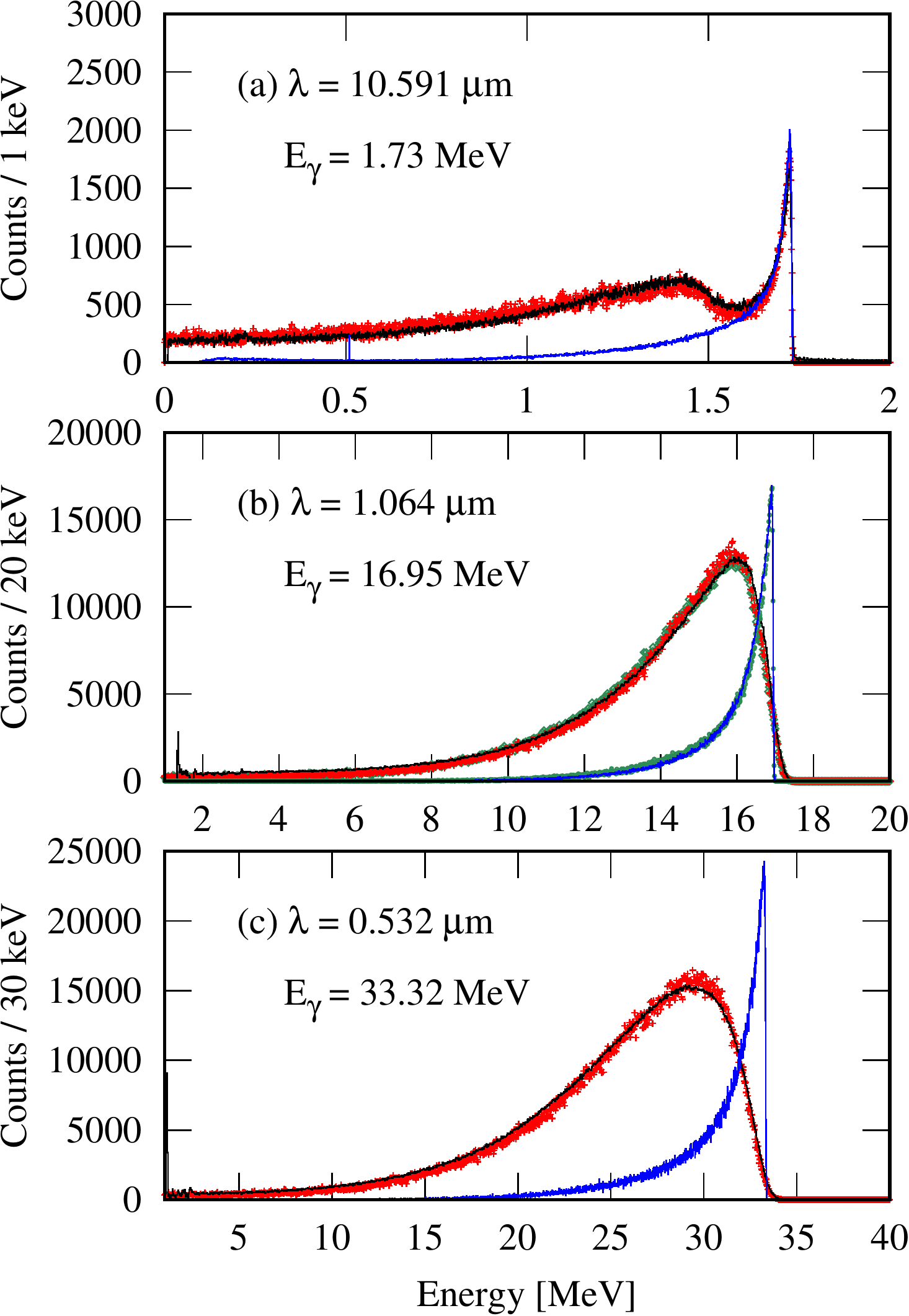} 
\caption{Experimental (black lines) and simulated (red pluses) response functions of (a) a 64.3~mm diameter $\times$ 60.1~mm length HPGe detector and of (b-c) a 3.5" diameter $\times$ 4" length LaBr$_3$:Ce detector to LCS $\gamma$-ray beams and the simulated energy spectra of the incident LCS $\gamma$-ray beams (blue lines). The LCS $\gamma$-ray beams were obtained using NewSUBARU electron beams at the injection energy of 982.43~MeV and (a) 10.591~$\mu$m, (b) 1.064~$\mu$m, and (c) 0.532~$\mu$m wavelength lasers. Two alignment settings are discussed in the text for the 1.064~$\mu$m laser. The energy spread in the full width at half maximum is of 1.85$\%$, 2.53$\%$, and 2.85$\%$ for 1.73, 16.95, and 33.32~MeV LCS $\gamma$-ray beams, respectively.      
}
\label{fig_TOP_UP}       
\end{figure}

\subsection{Scattered $\gamma$-ray photon transport }\label{label_sub_sub_sec_transport_g}

The sampling procedure is implemented in a dedicated \texttt{C++} class included in the \texttt{eliLaBr} \textsc{Geant4} simulation code and called by the \texttt{PrimaryGeneratorAction} class. 
Thus, the $\gamma$-ray photon generated at Sec.~\ref{label_sub_sub_sec_sample_g} and, if needed, the corresponding recoil electron are transported within the \textsc{Geant4} simulations. 
The general layout of the experimental setup follows the NewSUBARU experimental facility with all characteristic sizes, distances and materials: beamline BL01, synchrotron dipole magnetic field, laser optical mirror used to insert the laser beam into the electron beamline, borosilicate vacuum window, $\gamma$-beam shutter, primary and secondary $\gamma$-beam collimators, as well as detection systems placed in the experimental hutch. 

\subsection{Example of LCS $\gamma$-ray beam flux and energy simulations}
\label{label_subsec_sim_example}

Figure~\ref{fig_sim_spot} shows simulations of LCS $\gamma$-ray beam spectra and incident flux obtained for different configurations of the Talon laser while keeping the electron beam conditions fixed (top-up energy $E_e$~=~982.43~MeV and good transverse emittance values of $\varepsilon_x$~=~37~nm~rad, $\varepsilon_y$~=~3.7~nm~rad). The left column shows the overlap between the laser and electron beams along the interaction region, as defined by Eq.~\ref{label_eq_luminosity_approximate_general_expl}. 
The collimated $\gamma$-ray spectra are shown in absolute value in the middle column, while the right column gives the same spectra normalized in order to compare the bandwidth. A collimation configuration of C1~=~3~mm and C2~=~2~mm has been considered. 

Figure~\ref{fig_sim_spot}(a1--3) shows a study for different laser beam diameter values D4$\sigma$ at laser output, with the Talon nominal D4$\sigma$~=~1.048~mm. We notice in Figure~\ref{fig_sim_spot}(a1) that the distance between the interaction points and the collimator increases with the decrease in the D4$\sigma$ laser diameter. This leads to a narrower collimation cone and an improved energy resolution, as shown in Fig.~\ref{fig_sim_spot}(a2--3).  

Figure~\ref{fig_sim_spot}(b1--3) shows a study for vertical and horizontal laser 1~mm off-sets relative to the electron beam axis, which produce significant drops in the LCS $\gamma$-ray beam flux. We notice that the flux decrease induced by a vertical off-set is stronger than the one induced by an identical 1~mm horizontal off-set. 

In Figure~\ref{fig_sim_spot}(c1--3), different D4$\sigma$ values and a small $\theta$ rotation of the laser beam relative to the electron axis have been considered. It can be observed that the interactions mainly take place in the region of the laser focus, because, as can be seen from equations~\ref{eq_rotation} and \ref{eq_translation}, first the rotation of the laser is performed around the laser focus and then the entire laser is translated.

\begin{table}[t]
\begin{center}
\caption {Parameters for the NewSUBARU electron beam. Besides the nominal emittance values~\cite{aoki2004,miyamoto2007,amano09,Horikawa2010}, we also list the values resulted by reproducing the experimental detector response to LCS $\gamma$-ray beams shown in Fig.~\ref{fig_TOP_UP}.} 
\label{tab_electrons}
  \begin{tabular}[width=0.45\textwidth]{   l r   r  }
  \hline \hline
    Energy   & min. &   550 MeV  \\ 
             & max. & 1460 MeV  \\
             & injection $\&$ top-up &   982.43 MeV \\    
    rms $\Delta$E at top-up & & 0.04 \% \\ \hline   \hline 
    Emittance (nm-rad) & & ($\varepsilon_x$, $\varepsilon_y$)  \\  
    Nominal at top-up  &  & (38, 1~--~3.8)  \\    
    \multicolumn{2}{l}{Best fit values to LCS $\gamma$-ray data:} & \\ \hline
    CO$_2$  &  10.591  $\mu$m & (60, 6.)  \\
            &  $E_\gamma^{max}$= 1.73~MeV & \\
            &  (C1, C2) = (3, 1) mm       & \\
            &  HPGe monitor               & \\            \hline
    INAZUMA &  1.064  $\mu$m              & (55, 5.5)  \\
            &  $E_\gamma^{max}$= 16.95~MeV&  \\
            &  (C1, C2) = (6, 2) mm       & \\
            &  LaBr$_3$:Ce monitor        & \\ \hline
    Talon   &  0.532 $\mu$m               & (80, 8)  \\
            &  $E_\gamma^{max}$= 33.32~MeV &  \\
            &  (C1, C2) = (3, 2) mm       & \\
            &  LaBr$_3$:Ce monitor        & \\    \hline \hline           
  \end{tabular}
\end{center}
\end{table}

\section{Comparison with experimental LCS $\gamma$-ray beam spectra}
\label{label_comparison_with_exp}

\subsection{Energy distribution}
\label{label_sub_sec_cmp_exp_en_dist}

Fig.~\ref{fig_TOP_UP} shows experimental and simulated detector response functions to LCS $\gamma$-ray beams produced at NewSUBARU for different lasers, collimation settings and detectors. For all three cases exemplified here, the electron beam had the injection energy of 982.43~MeV with rms energy resolution of 0.04$\%$ and the laser beam has a 100$\%$ linear polarization in the vertical plane. The simulation results are summarized in Table~\ref{tab_electrons}. We note that, when reproducing the experimental LCS $\gamma$-ray spectra, a 10\% ratio was kept between the vertical and horizontal emittances, following the electron beam measurements reported in Refs.\cite{aoki2004,miyamoto2007,amano09,Horikawa2010}.

Fig.~\ref{fig_TOP_UP}(a) shows a 1.73~MeV maximum energy $\gamma$-ray beam produced using the 10.591~$\mu$m wavelength CO$_2$ laser. 
The LCS $\gamma$-ray beam irradiated a coaxial HPGe detector (64~mm diameter $\times$ 60 mm length) placed with a 15~mm offset on the $y$ axis relative to the $\gamma$-ray beam axis, in order to avoid direct irradiation of the central Ge crystal hole. 
The CO$_2$ laser is aligned with the electron beam axis and its nominal parameters given in Table~\ref{tab_lasers} have been considered. The best fit electron emittance values are listed in Table~\ref{tab_electrons}.  

\begin{figure}[t]
\centering
\includegraphics[width=0.45\textwidth, angle=0]{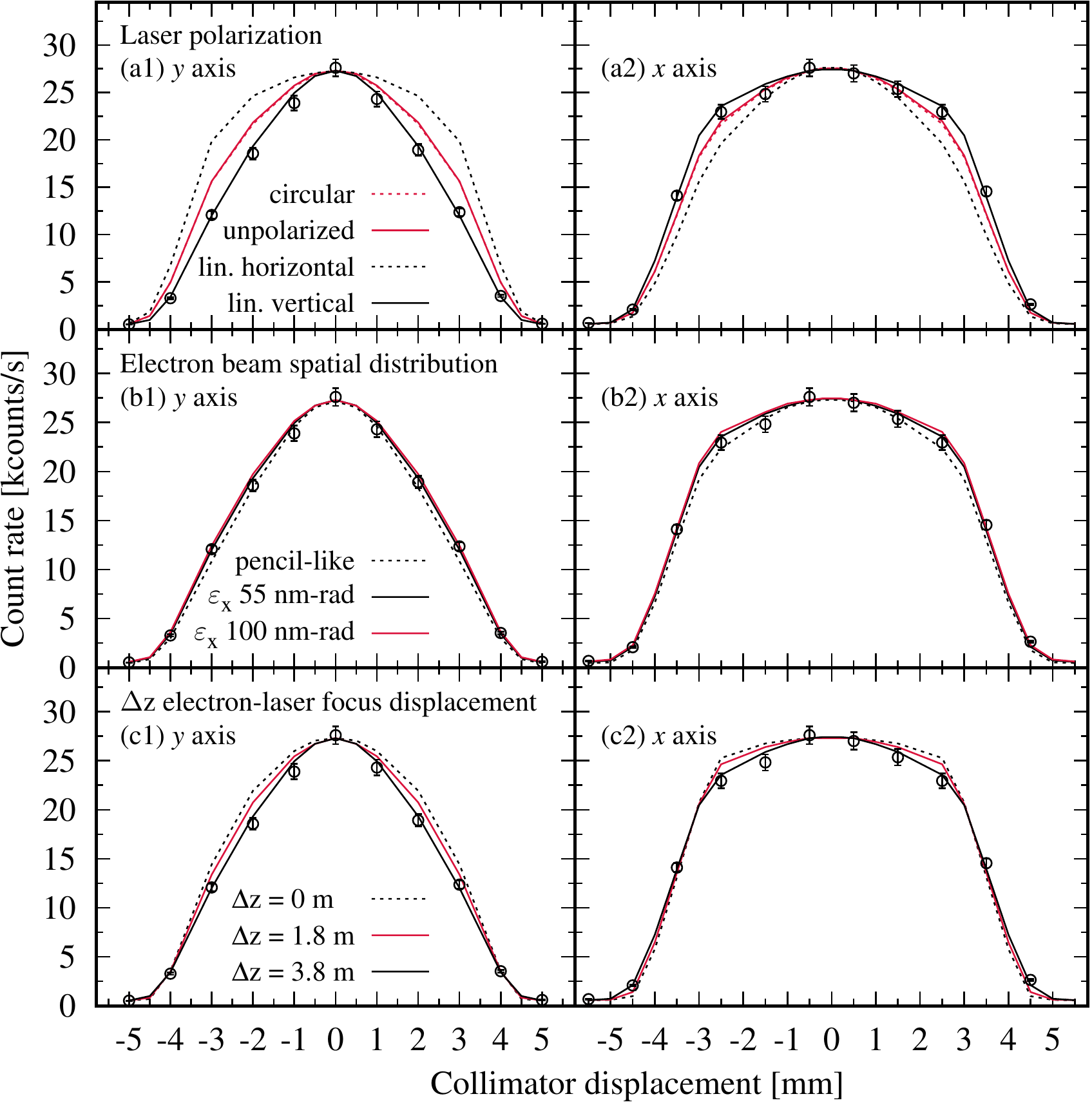}
\caption{Experimental flux values (circles) for 16.95~MeV energy $\gamma$-ray beams (982.54~MeV electrons, 1.064~$\mu$m laser) recorded by the 3.5" $\times$ 4" LaBr$_3$:Ce detector as a function of the 2~mm aperture C2 collimator displacements on the $y$-vertical (left column) and $x$-horizontal (right column) axis. Experimental values are compared with simulations, where we have investigated the effect brought to the $\gamma$-ray flux spatial distribution by the (a1-2) laser polarization, (b1-2) spatial distribution of the electron beam and (c1-2) the displacement along the $\Delta z$ axis between the electron beam and laser beam focus positions. 
}
\label{fig_col_area}       
\end{figure}

Fig.~\ref{fig_TOP_UP}(b) shows a 16.95~MeV maximum energy $\gamma$-ray beam produced using the 1.064~$\mu$m wavelength Nd:YVO$_4$ INAZUMA laser. 
LaBr$_3$:Ce detector response simulations are shown for aligned laser~--~electron beam conditions (green diamonds) with nominal INAZUMA laser parameters given in Table~\ref{tab_lasers}. We also show simulated detector response (red pluses) obtained by considering for the laser beam a different than the nominal focus position, a small horizontal offset of 0.1~mm relative to the electron beam axis and a 0.44~mrad rotation around the $y$ axis (in the horizontal plane). The incident spectrum obtained for aligned laser~--~electron beam conditions is shown by green circles and the one obtained including the small laser misalignment is shown in blue lines, with good agreement between the two spectra. 
Although the simulations for both laser configurations reproduce well the experimental detector response, the laser misalignment has been introduced based on an additional investigation of the spatial distribution of the $\gamma$-ray flux and energy spectrum which we detail in Sec.~\ref{subsec_col}. 

Fig.~\ref{fig_TOP_UP}(c) shows a 33.32~MeV maximum energy $\gamma$-ray beam produced using the 0.532~$\mu$m wavelength solid state Talon laser and the C1~=~3~mm and C2~=~2~mm collimator configuration. 
The Talon laser is aligned with the electron beam axis and its nominal parameters given in Table~\ref{tab_lasers} have been considered.  

\begin{figure*}[t]
\centering
\includegraphics[width=0.9\textwidth, angle=0]{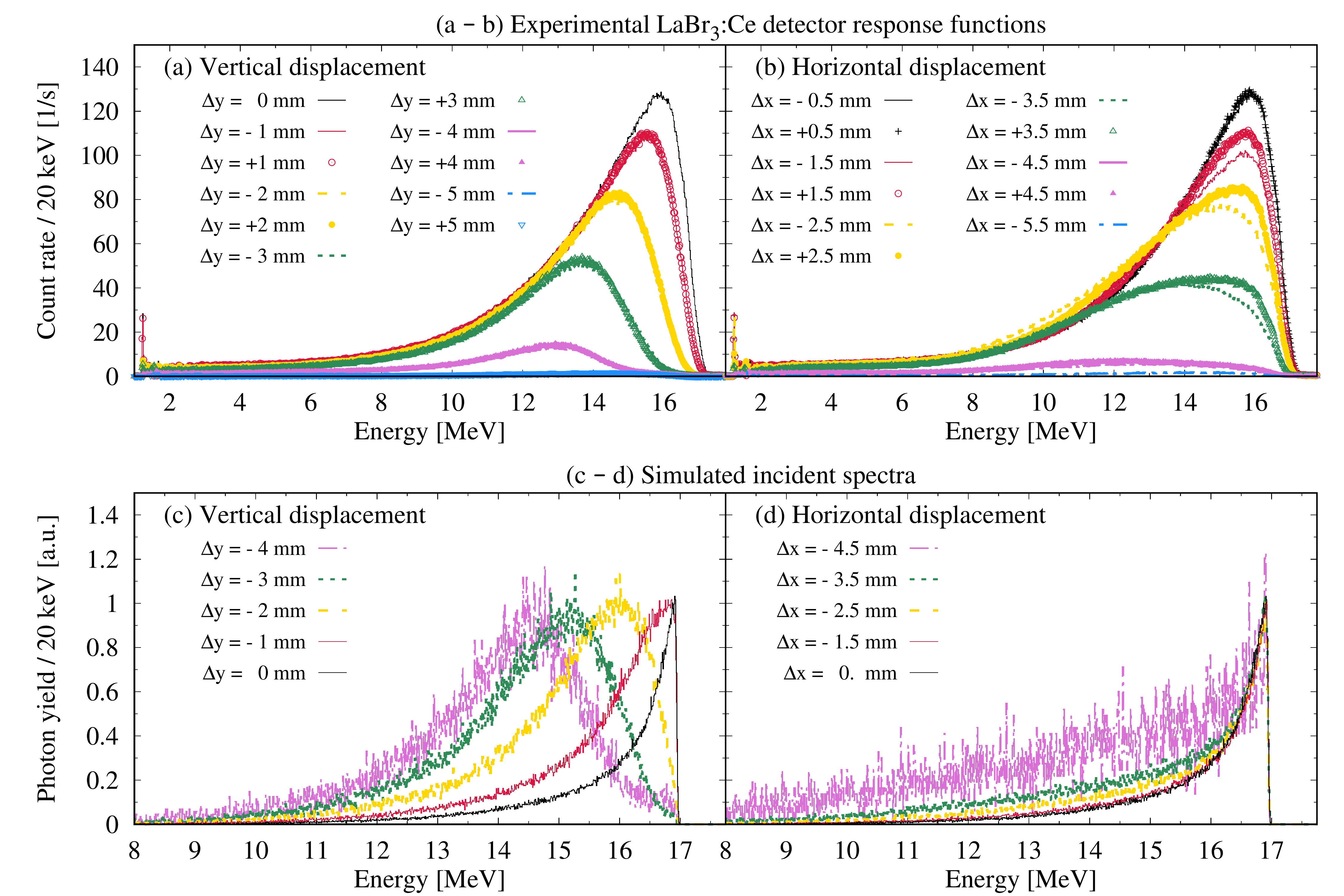}   
\caption{Effects of transverse collimator offsets on LCS $\gamma$-ray beams flux, spectral distribution and maximum energy. 16.95~MeV $\gamma$-ray beams were produced with 982.43~MeV electrons and a 1.064~$\mu$m laser. Two collimators of C1~=~6~mm and C2~=~2~mm aperture were used. 
Experimental response functions of a 3.5" $\times$ 4" LaBr$_3$:Ce detector are shown for different (a) vertical ($\Delta$y) and (b) horizontal ($\Delta$x) displacements of the C2 collimator relative to the $\gamma$-ray beam axis. Simulated incident energy spectra are shown for similar (c) vertical and (d) horizontal offsets. The incident spectra are scaled to their respective peak values. We note the high energy edge degradation for increasing vertical offsets. 
}
\label{fig_col_exp}       
\end{figure*}

\subsection{Spatial distribution of $\gamma$-ray flux and energy spectrum}
\label{subsec_col}

We here benchmark our simulation code against experimental investigations of spatial distribution of LCS $\gamma$-ray flux and energy spectrum. For the experimental investigation we have used the 1.064~$\mu$m wavelength Nd:YVO$_4$ laser and electron beams at the injection energy of 982.43~MeV at constant 300~mA current. A double collimation system of C1~=~6~mm and C2~=~2~mm was used. Prior to the measurement, both collimators have been aligned with the electron and laser beam axis by maximizing the LCS $\gamma$-ray flux. With the coarse C1~=~6~mm collimator in the optimum position, the energy spectra and flux of the LCS $\gamma$-ray beams have been investigated for horizontal and vertical displacements of the fine C2~=~2~mm collimator. The experimental and simulated LaBr$_3$:Ce detector response functions as well as the incident spectra for the central position have been shown in Fig.~\ref{fig_TOP_UP}(b) and discussed in Sec.~\ref{label_sub_sec_cmp_exp_en_dist}.  

\subsubsection{Spatial distribution of $\gamma$-ray flux}

Fig.~\ref{fig_col_area} shows the LCS $\gamma$-ray flux recorded by the 3.5" $\times$ 4" LaBr$_3$:Ce detector as a function of the vertical (left column) and horizontal (right column) displacement of the C2 collimator relative to the electron and laser beam axis. The experimental results (circles) show asymmetric behaviors on the transverse planes, with a narrow flux distribution on the vertical axis and a rather flat top distribution on the horizontal axis. 

The laser polarization effect on the spatial distribution of the $\gamma$-ray flux is investigated in Fig.~\ref{fig_col_area}(a1,a2), where unpolarized electron beams were considered. We observe that the experimental results are well described by considering laser beams linearly polarized in the vertical plane, which is the polarization configuration of the INAZUMA laser. 

The electron beam phase-space distribution effect on the spatial distribution of the $\gamma$-ray flux is investigated in Fig.~\ref{fig_col_area}(b1,b2). We simulated $\gamma$-ray flux distribution curves for the best-fit emittance parameters ($\varepsilon_x$,~$\varepsilon_y$)~=~(55,~5.5)~nm-rad obtained at Sec.~\ref{label_sub_sec_cmp_exp_en_dist} and for significantly worse emittance values of ($\varepsilon_x$,~$\varepsilon_y$)~=~(100,~10)~nm-rad, as well as for ideal pencil-like electron beams. Figure~\ref{fig_col_area}(b1,b2) shows that the spatial distribution of the $\gamma$-ray flux on both the vertical and horizontal axis is insensitive to the electron beam spatial distribution.  

The geometrical effect given by the position of the interaction point along the interaction line is investigated in Fig.~\ref{fig_col_area}(c1,c2). Having the electron beam focus fixed in the center of the interaction line, we considered three positions for the laser beam focus: in the electron beam focus with a corresponding displacement $\Delta z$~=~0~m (black dotted line), at the nominal $\Delta z$~=~1.8~m displacement towards the collimation system (full red line) and at an increased $\Delta z$~=~3.8~m displacement (full black line). We find that the $\gamma$-ray flux spatial distribution narrows with the decrease in the distance between the effective interaction region and the collimation system. The experimental data are best described by simulations performed with the $\Delta z$~=~3.8~m displacement, the same value used also for the simulations shown in Figs.~\ref{fig_col_area}(a,b). 

\begin{figure*}[t]
\centering
\includegraphics[width=0.9\textwidth, angle=0]{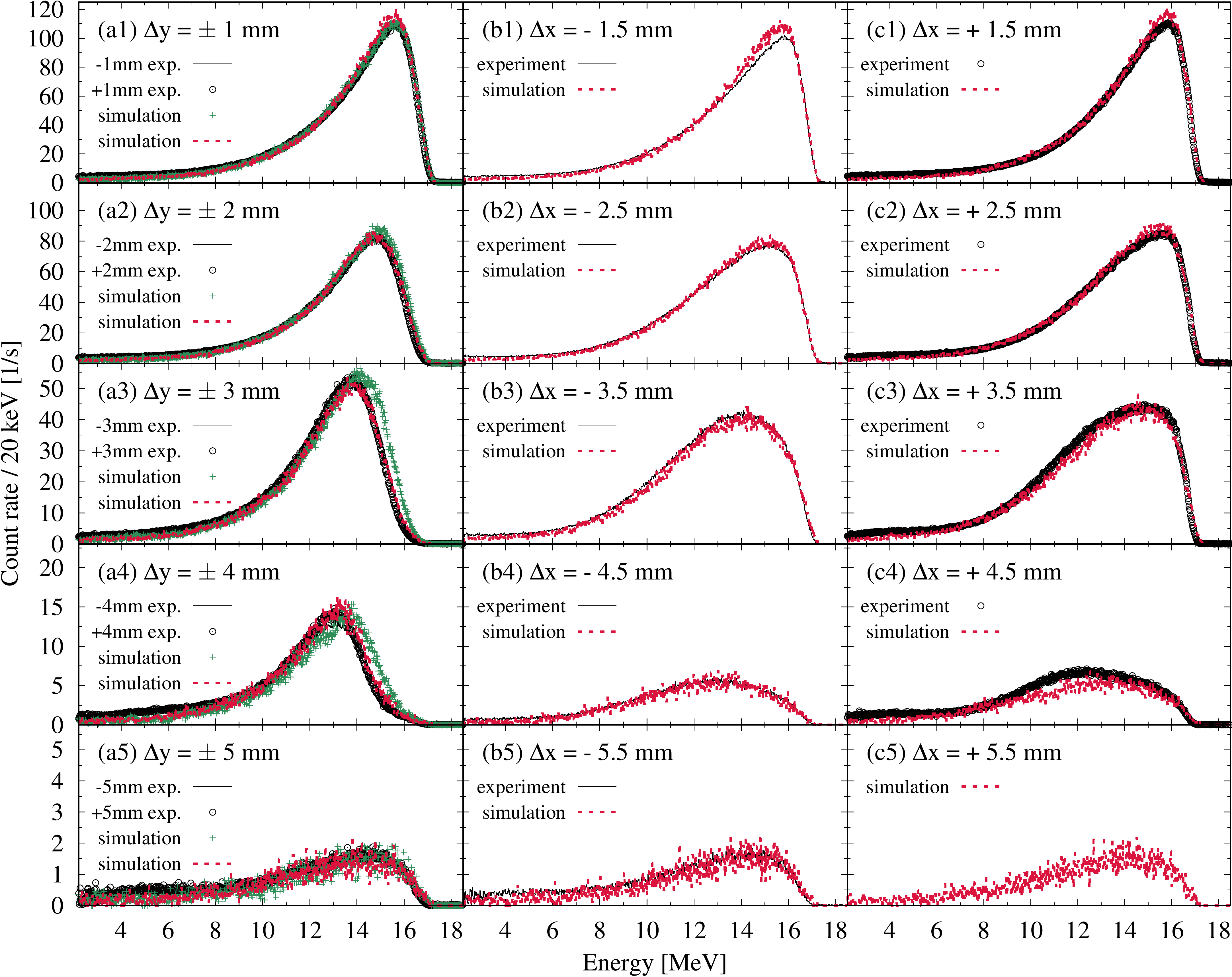}
\caption{Experimental and simulated response functions of a 3.5" diameter $\times$ 4" length LaBr$_3$:Ce detector to 16.95~MeV LCS $\gamma$-ray beams for different (a1-5) vertical ($\Delta$y) and (b1-5,c1-5) horizontal ($\Delta$x) displacements of the C2~=~2~mm aperture collimator relative to the electron beam axis. All simulations are performed with $\varepsilon_x$~=~55~nm-rad and $\varepsilon_y$~=~5.5~nm-rad emittance electron beams and laser beams linearly polarized in the vertical plane. Simulations performed with the INAZUMA laser aligned with the electron beam and at the nominal $\Delta$z~=~1.8~m displacement between the focus of the electron and laser beams are given in green pluses. Simulations performed with the laser rotated by $\theta$~=~0.44~mrad, horizontally displaced by 0.1~mm relative to the electron axis and with a $\Delta$z~=~3.8~m displacement between the focus of the electron and laser beams are shown in dashed red line.          
}
\label{fig_col_exp_sim}       
\end{figure*}

\subsubsection{Spatial distribution of $\gamma$-ray energy spectrum}\label{label_sub_sub_sec_spatial_distr_en_sp}

\begin{figure*}[t]
\centering
\includegraphics[width=0.9\textwidth, angle=0]{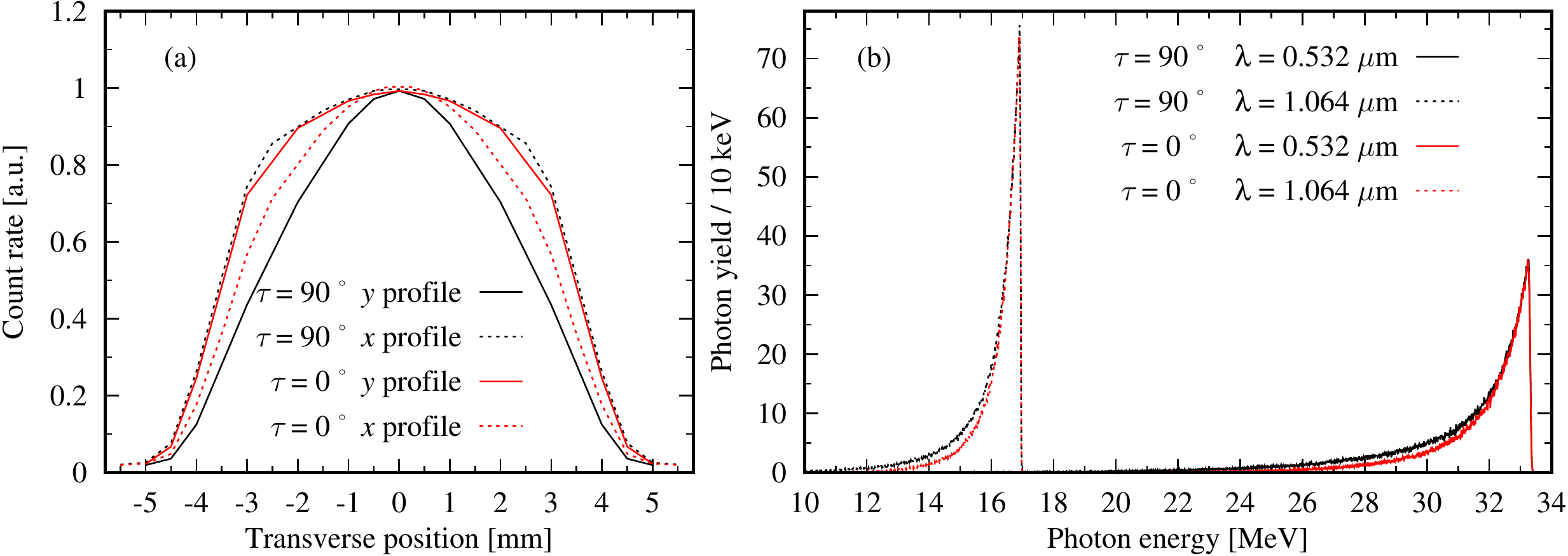}   
\caption{Effects of the laser polarization plane orientation ($\tau$) on the collimated LCS $\gamma$-ray beam flux spatial distribution and energy spectrum. We consider laser beams 100~\% linearly polarized along the $y$ axis (polarization angle $\tau$~=~90$^\circ$ -- perpendicular to the accelerator plane -- black lines) and along the $x$ axis ($\tau$~=~0$^\circ$ -- in the accelerator plane -- red lines). (a) Flux spatial distribution along the $y$ (full lines) and $x$ (dotted lines) axes. (b) Energy spectra of LCS $\gamma$-ray beams obtained with $E_e$~=~982.43~MeV electron beams and 0.532~$\mu$m (full lines) and 1.064~$\mu$m (dotted lines) lasers. 
}
\label{fig_11_test_area3}       
\end{figure*}

Fig.~\ref{fig_col_exp} shows the effects of the transverse collimator offsets on the spectral distribution, maximum energy and flux of the 16.95~MeV LCS $\gamma$-ray beams. Experimental LaBr$_3$:Ce response functions are shown in the upper row of figures, and normalized simulated incident energy spectra are shown in the lower figures. The well known~\cite{Sun2009_STAB,Horikawa2010} incident flux decrease with the increase in the collimator transverse displacement is observed in the areas of the experimental spectra and in the fluctuations of the simulated ones, for both the vertical and horizontal planes. However, we observe an asymmetry in the transverse spatial distribution of the $\gamma$-ray energy spectrum.

A vertical profile investigation is shown in figures~\ref{fig_col_exp}(a) and (c). Figure~\ref{fig_col_exp}(c) shows that the Compton backscattered, high energy region of the spectral distributions is increasingly cut-away with the increase in the vertical collimator offsets. Because of the non-unity detection efficiency and characteristic energy resolution of the 3.5" $\times$ 4" LaBr$_3$:Ce detector, an apparent decrease of the maximum energy edge of $\sim$0.5~MeV per mm of vertical displacement is observed in the experimental response functions shown in figure~\ref{fig_col_exp}(a).

Fig.\ref{fig_col_exp}(b) and (d) show the spatial distribution of $\gamma$-ray energy spectrum in the horizontal plane. While the energy resolution of the spectral distributions is worsening with the increase in the horizontal collimator displacement, the maximum energy edge is not altered. The results reproduce the calculations of Sun {\it et al.}~\cite{Sun2009_STAB} for small offsets compared to the collimation aperture size. Unlike the up/down symmetrical vertical profiles, the spectra recorded for equal horizontal displacements show a left/right asymmetry with sharper maximum energy front lines for positive $\Delta x$ displacements. 

Fig.\ref{fig_col_exp_sim} shows the experimental detector response functions for vertical and horizontal collimator displacements reproduced by simulations. For the simulations, we considered the transverse emittance parameters of $\varepsilon_x$~=~55~nm-rad and $\varepsilon_y$~=~5.5~nm-rad found to reproduce well the LaBr$_3$:Ce detector response recorded in aligned collimator conditions shown in Fig.~\ref{fig_TOP_UP}(b). Laser beams linearly polarized along the vertical plane have been considered, accordingly to the INAZUMA laser specifications and based on the good reproduction of the spatial distribution of the $\gamma$-ray flux shown in Fig.~\ref{fig_col_area}(a1,a2). Based on the flux distribution study given in the previous section, two configurations have been considered for the displacement between the INAZUMA laser and electron beam focal positions: the nominal value of $\Delta z$~=~1.8~m and the increased $\Delta z$~=~3.8~m.  

The experimental detector response functions recorded for vertical collimator displacements are compared in Fig.\ref{fig_col_exp_sim}(a1-5) with simulation results obtained with the $\Delta z$~=~1.8~m (green pluses) and with $\Delta z$~=~3.8~m (dashed red line) laser displacement values. As is the case for the un-shifted collimator position shown in Fig.~\ref{fig_TOP_UP}(b), simulations performed with both laser displacement configurations reproduce well the experimental spectra obtained with small vertical collimator offsets of $\pm$~1~mm and $\pm$~2~mm shown in Figs.\ref{fig_col_exp_sim}(a1,a2). However, Figs.\ref{fig_col_exp_sim}(a3,a4) show that the experimental results obtained with the larger $\pm$~3~mm and $\pm$~4~mm vertical collimator offsets are well reproduced only by the $\Delta z$~=~3.8~m configuration results, while the simulations performed with the nominal $\Delta z$~=~1.8~m INAZUMA laser parameters overestimate the experimental maximum energy front. 
 
In order to further reproduce the left/right asymmetry observed in the experimental detector response functions recorded for horizontal displacements of the collimators, a small angular misalignment of 0.44~mrad around the $y$ axis and a 0.1~mm horizontal offset have been applied to the laser beam. Fig.\ref{fig_col_exp_sim}(b1-5) and Fig.\ref{fig_col_exp_sim}(c1-5) show that the so obtained simulated results reproduce well the experimental detector response functions recorded both for horizontal and respectively vertical displacements of the collimator.  

\section{Laser polarization orientation for LCS $\gamma$-ray beams produced with synchrotron electron beams} 
\label{label_sec_laser_pol}

Considering the characteristic asymmetry between the vertical and horizontal emittance profiles for synchrotron electron beams, we find it most relevant to study the case of 100\% linearly polarized laser beams with polarization vector in the accelerator plane -- along the horizontal $x$ axis, and perpendicular to the accelerator plane -- along the vertical $y$ axis. We denote the horizontal orientation by polarization angle $\tau$~=~0$^\circ$ and the vertical one by $\tau$~=~90$^\circ$. For the two polarization configurations, we will focus our investigations on the two critical aspects emphasized in Section~\ref{label_sec_motivation}:
\begin{itemize}
\item preservation of LCS $\gamma$-ray beam maximum energy edge;
\item good characterization of the LCS $\gamma$-ray spectra.
\end{itemize}

\subsection{Maximum energy edge preservation}
\label{subsub_pol_max_en}

We showed in Sec.~\ref{label_sub_sub_sec_spatial_distr_en_sp} that the maximum energy of the LCS $\gamma$-ray beam is influenced by the offsets of the collimators on the vertical axis. As the alignment of the collimator to the laser and electron beam axis is based on maximizing the LCS $\gamma$-ray flux, the alignment precision is determined by the variation in the spatial distribution of the $\gamma$-ray beam intensity. 

Figure~\ref{fig_11_test_area3}(a) shows the vertical and horizontal LCS $\gamma$-ray beam intensity profiles for incident laser beams 100$\%$ linearly polarized along the vertical and horizontal axes. As expected~\cite{Sun2011_STAB,petrillo2015,Hajima2021,paterno_2022}, the transverse spatial distribution follows the laser polarization orientation: $\gamma$-rays scatter predominantly along the horizontal axis for vertical laser polarization, and along the vertical axis for horizontal laser polarization.

However, the switch between the horizontal / vertical profiles obtained by flipping the laser polarization orientation between the horizontal / vertical axis is not symmetric: 
\begin{itemize}
\item for vertically polarized laser ($\tau$~=~90$^\circ$), the beam intensity along the vertical axis shows a strong sensitivity to small offsets in the vicinity of the central position, while the horizontal profile is rather flat-top;  
\item for horizontally polarized laser ($\tau$~=~0$^\circ$), there is less difference between the horizontal and vertical profiles, both being wider than the vertical profile for $\tau$~=~90$^\circ$. Considering the asymmetrical spatial distribution of the electron beams, this suggests a smearing effect introduced by the larger horizontal emittance and thus rms electron beam distribution.
\end{itemize}
Therefore, the precise vertical collimator alignments necessary for preserving the maximum LCS $\gamma$-ray energy are facilitated by using vertical polarized lasers.

\subsection{LCS $\gamma$-ray spectral distributions}
\label{subsub_pol_lcs_sp}

Figure~\ref{fig_11_test_area3} shows simulations for LCS $\gamma$-ray spectra obtained with $E_e$~=~982.43~MeV electron beams and 1.064~$\mu$m INAZUMA and 0.532~$\mu$m Talon lasers. The laser~--~electron interaction takes place in head-on conditions for perfectly aligned beams. A collimation configuration of C1~=~3~mm and C2~=~2~mm has been considered. Good electron beam conditions were considered, of $\varepsilon_x$~=~38~nm-rad horizontal and $\varepsilon_y$~=~3.8~nm-rad vertical emittance. 

We notice that, for both investigated lasers, the LCS $\gamma$-ray spectra improve for horizontal laser polarization. 
The spectra obtained using lasers polarized in the vertical plane show a tail towards low energies more amplified than in the case of horizontal polarization.
The ratio between the integrated spectra for the horizontal and vertical laser polarization orientations is of $\sim$80\% for Talon and $\sim$85\% for INAZUMA. The ratio remains approximately constant also for narrower collimation conditions. 

\subsection{Laser polarization configuration for GDR photonuclear experiments}

In Section~\ref{subsub_pol_max_en} we showed that the precision in the vertical alignment of the collimator is increased by using vertically polarized lasers in connection with electron beams of asymmetric emittance profiles. However, as shown in Section~\ref{subsub_pol_lcs_sp}, by using vertically polarized lasers, one amplifies the low energy tail in the spectral distribution of LCS $\gamma$-ray beams. Thus, we conclude that the preservation of the maximum energy edge through good vertical collimator alignments with vertically polarized lasers comes at the price of energy resolution deterioration, as summarized in Table~\ref{tab_results}.  

\section{Conclusion}
\label{label_sec_conclusions}

Following an extensive campaign of photoneutron measurements at the NewSUBARU LCS $\gamma$-ray beamline, 
we have developed the \texttt{eliLaBr} Monte Carlo simulation code for characterization of the scattered $\gamma$-ray photon beams. The code is implemented using \textsc{Geant4} and is available on the GitHub repository~\cite{eliLaBr_github}. 
The present work treats the spectral distribution and flux of LCS $\gamma$-ray beams with focus on the realistic modeling of continuous,  unsynchronized laser and relativistic electron beams.  

The results of the \texttt{eliLaBr} code were validated against NewSUBARU experimental data of LCS $\gamma$-ray flux and energy spectra spatial distributions taken in aligned conditions and also for a systematic investigation of transverse collimator offsets relative to the laser and electron axis. We have shown that the maximum energy of the LCS $\gamma$-ray beam is altered by vertical collimator offsets, where the edge shifts towards lower energies with the increase of the vertical collimator offset. However, horizontal collimator offsets do not alter the maximum energy edge, but lead to the well-known flux decrease\cite{Horikawa2010} and energy resolution deterioration~\cite{Sun2009_STAB}.

\begin{table}[t]
\begin{center}
\caption {Main results of laser polarization orientation study.} 
\label{tab_results}
  \begin{tabular}[width=0.45\textwidth]{   l r   }
  \hline \hline
    Laser polarization & Advantage  \\ \hline \hline
    Vertical           & Preservation of maximum energy   \\
                       & edge through precise vertical alignment \\ \hline
    Horizontal         & Improved resolution of    \\  
                       & LCS $\gamma$-ray beam spectral distributions \\ \hline \hline           
  \end{tabular}
\end{center}
\end{table}

Considering the asymmetric transverse emittance profiles characteristic for synchrotron electron beams, we have investigated the effect of the laser polarization plane orientation on the properties of the LCS $\gamma$-ray beams. We show that:
\begin{itemize}
\item the use of vertically polarized lasers increases the precision in the vertical collimator alignment, leading thus to the preservation of the maximum energy edge; 
\item the use of horizontally polarized lasers improves the resolution of the LCS $\gamma$-ray beam spectral distribution.
\end{itemize}

For photonuclear experiments performed at LCS $\gamma$-ray energies of tens of MeV, knowing the absolute value of the maximum photon energy is critical, as it cannot be measured experimentally. On the other hand, through a good modeling of the LCS $\gamma$-ray beam, the energy spectrum and the resolution can be precisely reproduced and treated accordingly in the data analysis, as we detailed in Sec.~\ref{label_sec_motivation}. This results in an advantage of using vertically polarized lasers for measurements with LCS $\gamma$-ray beams performed at electron synchrotrons.

The present paper follows our recent Monte Carlo study~\cite{Filipescu_2022_POL} on the polarization of the Compton scattered photon. Experimental and simulation investigations of the spatial distribution of the LCS $\gamma$-ray beam at target position will be reported in a follow-up paper~\cite{takashi_minipix}.

\section*{Acknowledgments}
\label{label_SEC_acknowledgments}

The authors are grateful to Octavian Sima of the University of Bucharest for correspondence and useful discussions on the present modeling of laser and electron beam interaction. 
D.F. acknowledges the support from the Romanian Ministry of Research, Innovation and Digitalization/Institute of Atomic Physics from the National Research - Development and Innovation Plan III for 2015--2020/Programme 5/Subprogramme 5.1 ELI-RO, project GANT-Photofiss No 14/16.10.2020. 
This work was supported by a grant of the Ministry of Research, Innovation and Digitization, CNCS - UEFISCDI, project number PN-III-P1-1.1-PD-2021-0468, within PNCDI III.


\begin{thebibliography}{00}



\bibitem{Shizuma2021} T.~Shizuma {\it et al.}, Low-lying electric and magnetic dipole strengths in $^{207}$Pb, Phys. Rev. C \textbf{103}, 024309 (2021).
\bibitem{KEIde2021} K.E.~Ide {\it et al.}, $E2$ decay characteristics of the $M1$ scissors mode of $^{152}$Sm, Phys. Rev. C \textbf{103}, 054392 (2021).
\bibitem{Weller2009} Henry~R.~Weller {\it et al.}, Research opportunities at the upgraded HI$\gamma$S facility, Progress in Particle and Nuclear Physics, \textbf{62}, 257 (2009).
\bibitem{WLuo2016} Wen~Luo {\it et al.}, Estimates for production of radioisotopes of medical interest at Extreme Light Infrastructure – Nuclear Physics facility, Appl. Phys. B \textbf{122}, 8 (2016). 
\bibitem{PGros2018} P.~Gros {\it et al.}, Performance measurement of HARPO: A time projection chamber as a gamma-ray telescope and polarimeter, Astroparticle Physics \textbf{97}, 10 (2018). 
\bibitem{HLan2021} Haoyang~Lan {\it et al.}, Nuclear resonance fluorescence drug inspection, Nature Scientific Reports \textbf{11}, 1306 (2021).
\bibitem{Klein2002} R.~Klein {\it et al.}, Measurement of the BESSY II electron beam energy by Compton-backscattering of laser photons, Nucl. Instrum. Methods Phys. Res. A {\bf 486}, 545 (2002).
\bibitem{Sun2009_STAB} C.~Sun {\it et al.}, Energy and energy spread measurements of an electron beam by Compton scattering method, Physical Review Special Topics - Accelerators and Beams \textbf{12}, 062801 (2009).
\bibitem{Chouffani2006} K.~Chouffani {\it et al.}, Determination of electron beam parameters by means of laser-Compton scattering, Physical Review Special Topics - Accelerators and Beams \textbf{9}, 050701 (2006).
\bibitem{litvinenko_1997} V.N.~Litvinenko et al., Gamma-ray production in a storage ring free-electron laser, Phys. Rev. Lett. {\bf 78} (1997) 4569--4572. 
\bibitem{amano09}  S. Amano {\it et al.}, Several-MeV $\gamma$-ray generation  at NewSUBARU by laser Compton backscattering, Nucl. Instrum. Methods Phys. Res. A \textbf{602}, 337 (2009).
\bibitem{Horikawa2010} Ken~Horikawa {\it et al.}, Measurements for the energy and flux of Compton scattering $\gamma$-ray photons generated in an electron storage ring: NewSUBARU, Nucl. Instrum. Methods Phys. Res. A \textbf{618}, 209~--~215 (2010). 
\bibitem{muramatsu2022} N.~Muramatsu {\it et al.}, SPring-8 LEPS2 beamline: A facility to produce a multi-GeV photon beam via laser Compton scattering, Nucl. Instrum. Methods Phys. Res. A \textbf{1033} 166677 (2022). 
\bibitem{wang_fan_2022} Hong-Wei~Wang, Gong-Tao~Fan et al., Commissioning of laser electron gamma beamline SLEGS at SSRF, Nucl. Sci. Tech. {\bf 33}, 87 (2022). 
\bibitem{ZPan2019} Zhilong~Pan {\it et al.}, Design and dynamic studies for a compact storage ring to generate gamma-ray light source based on Compton backscattering technique, Physical Review Accelerators and Beams \textbf{22}, 040702 (2019).
\bibitem{Micieli2016} D.~Micieli {\it et al.}, Compton sources for the observation of elastic photon-photon scattering events, Physical Review Accelerators and Beams \textbf{19}, 093401 (2016).
\bibitem{facetII} Technical Design Report for the FACET-II Project at SLAC National Accelerator Laboratory. United States. https://doi.org/10.2172/1340171 
\bibitem{elinp_web} https://www.eli-np.ro/rd2$\_$second.php
\bibitem{HZen2016} Heishun~Zen {\it et al.}, Generation of High Energy Gamma-ray by Laser Compton Scattering of 1.94~$\mu$m Fiber Laser in UVSOR-III Electron Storage Ring, Energy Procedia \textbf{89}, 335 (2016).
\bibitem{Angelo2000} A.~D'Angelo {\it et al.}, Generation of Compton backscattering $\gamma$-ray beams, Nucl. Instrum. Methods Phys. Res. A {\bf 455}, 1 (2000). 
\bibitem{Petrillo2012} V.~Petrillo {\it et al.}, Photon flux and spectrum of $\gamma$-rays Compton sources, Nucl. Instrum. Methods Phys. Res. A {\bf 693}, 109 (2012). 
\bibitem{krafft2016} G. A. Krafft {\it et al.}, Laser pulsing in linear Compton scattering, Phys. Rev. Accel. Beams {\bf 19}, 121302 (2016).
\bibitem{CAIN} CAIN user manual, https://www-jlc.kek.jp/subg/ir/lib/cain21b.manual/ 
\bibitem{Sun2011_STAB} C.~Sun and Y.K.~Wu, Theoretical and simulation studies of characteristics of a Compton light source, Physical Review Special Topics - Accelerators and Beams \textbf{14}, 044701 (2011).  
\bibitem{Luo2011} W.~Luo {\it et al.}, A 4D Monte Carlo laser-compton scattering simulation code for the characterization of the future energy-tunable SLEGS, Nucl. Instrum. Methods Phys. Res. A \textbf{660} 108–115 (2011). 
\bibitem{Curatolo_PhD_thesis} C.~Curatolo, Ph.D. thesis, Universit\`{a} degli Studi di Milano, 2016, https://air.unimi.it/handle/2434/358227.
\bibitem{Curatolo2017} C.~Curatolo {\it et al.}, Analytical description of photon beam phase space in inverse Compton scattering sources, Physical Review Accelerators and Beams \textbf{20}, 080701 (2017). 
\bibitem{Hajima2021} Ryoichi~Hajima, Bandwidth of a Compton radiation source with an electron beam of asymmetric emittance, Nucl. Instrum. Methods Phys. Res. A \textbf{985} 164655 (2021). 
\bibitem{paterno_2022} Gianfranco Patern\`o {\it et al.}, Generation of primary photons through inverse Compton scattering using a Monte Carlo simulation code, Physical Review Accelerators and Beams \textbf{25}, 084601 (2022). 
\bibitem{Filipescu_2022_POL} D.~Filipescu, Monte Carlo simulation method of polarization effects in Laser Compton Scattering on relativistic electrons, JINST {\bf 17} P11006 (2022). 
\bibitem{geant_agostinelli_2003} S.~Agostinelli {\it et al.}, Geant4--a simulation toolkit, Nucl. Instrum. Methods Phys. Res. A, {\bf 506}, 250 (2003).   
\bibitem{geant_allison_2006} J.~Allison {\it et al.}, Geant4 developments and applications, IEEE Trans. Nucl. Sc. {\bf 53}, 270 (2006). 
\bibitem{geant_allison_2016} J.~Allison {\it et al.}, Recent developments in Geant4, Nucl. Instrum. Methods Phys. Res. A, {\bf 835}, 186 (2016).    
\bibitem{eliLaBr_github} D.~Filipescu and I.~Gheorghe, \texttt{eliLaBr} -- GEANT4 simulation code for LCS gamma-ray sources and flat efficiency moderated He-3 counters array dedicated to photoneutron reaction studies, https://github.com/dan-mihai-filipescu/eliLaBr (2022).
\bibitem{Gheorghe2017} I.~Gheorghe  {\it et al.}, Photoneutron cross-section measurements in the $^{209}$Bi($\gamma,\,xn$) reaction with a new method of direct neutron-multiplicity sorting, Phys. Rev. C \textbf{96}, 044604 (2017), and Phys. Rev. C \textbf{99}, 059901(E) (2019). 
\bibitem{Kawano2020} T. Kawano {\it et al.}, IAEA Photonuclear Data Library 2019,  Nucl. Data Sheets {\bf 163}, 109 (2020).
\bibitem{Filipescu_ND2022} D. Filipescu {\it et al.}, Photofission and photoneutron cross sections for $^{238}$U and $^{232}$Th, Contribution to ND2022, submitted. 
\bibitem{utsunomiyaNimDNM} H.~Utsunomiya {\it et al.}, Direct neutron-multiplicity sorting with a flat-efficiency detector, Nucl. Instrum. Methods Phys. Res. A {\bf 871}, 135--141 (2017).
\bibitem{IGheorghe_2021_MF} I.~Gheorghe {\it et al.}, Updated neutron-multiplicity sorting method for producing photoneutron average energies and resolving multiple firing events, Nucl. Instrum. Methods Phys. Res. A {\bf 1019}, 165867 (2021).
\bibitem{IEEE_Utsunomiya14} H.~Utsunomiya {\it et al.}, Energy Calibration of the NewSUBARU Storage Ring for Laser Compton-Scattering Gamma Rays and Applications, IEEE Transaction on Nuclear Science {\bf 61}, 1252 (2014).
\bibitem{Shima14} T.~Shima and H.~Utsunomiya, Energy Calibration of Electron and Gamma-Ray Beams at NewSUBARU-GACKO, in Proceedings of the Nuclear Physics and Gamma-ray Sources for Nuclear Security and Nonproliferation, January 28 - 30, 2014, Tokai, Japan, ed. T. Hayakawa {\it et al.}, World Scientific Publishing, Singapore, 151~--~160.
\bibitem{Toyokawa2000} Hiroyuki~Toyokawa {\it et al.}, Flux Measurement of the Laser-Compton-Backscattered Photons With a Poisson Fitting Method, IEEE Transaction on Nuclear Science {\bf 47}, 1954 (2000). 
\bibitem{kondo11} T. Kondo $et$ $al.$, Determination of the number of pulsed laser-Compton scattering photons, Nucl. Instrum. Methods Phys. Res. A \textbf{659}, 462~--~466 (2011).
\bibitem{utsunomiya_nimMP_2018} H. Utsunomiya {\it et al.}, Photon-flux determination by the Poisson-fitting technique with quenching corrections, Nucl. Instrum. Methods Phys. Res. A \textbf{896}, 103~--~107 (2018).
\bibitem{filipescu_2014_sm} D. Filipescu {\it et al.}, Photoneutron cross sections for samarium isotopes: Toward a unified understanding of ($\gamma$,~n) and (n,~$\gamma$) reactions in the rare earth region, Phys. Rev. C {\bf 90}, 064616 (2014).
\bibitem{gosta_2018} G.~Gosta {\it et al.}, Response function and linearity for high energy $\gamma$-rays in large volume LaBr$_3$:Ce detectors, Nucl. Instrum. Methods Phys. Res. A \textbf{879}, 92~--~100 (2018).
\bibitem{TRenstrom_unfolding} T.~Renstr\"om {\it et al.}, Verification of the detailed balance for $\gamma$ absorption and emission in Dy isotopes, Phys. Rev. C {\bf 98}, 054310 (2018).
\bibitem{aoki2004} K.~Aoki {\it et al.}, High-energy photon beam production with laser-Compton backscattering, Nucl. Instrum. Methods Phys. Res. A {\bf 516}, 228--236 (2004). 
\bibitem{miyamoto2007} Shuji Miyamoto {\it et al.}, Laser Compton back-scattering gamma-ray beamline on NewSUBARU, Radiation Measurements {\bf 41}, S179--S185 (2007).
\bibitem{utsunomiya_2015_npn} Hiroaki Utsunomiya {\it et al.}, The $\gamma$-Ray Beam-Line at NewSUBARU, Nuclear Physics News {\bf 25}, 25 -- 29 (2015).
\bibitem{mad_x} http://mad.web.cern.ch/mad/
\bibitem{WernerMuratori2006} Werner Herr and Bruno Muratori, Concept of luminosity, in CAS - CERN Accelerator School: Intermediate Course on Accelerator Physics, pp.361-378, 15 - 26 Sep 2003,  Zeuthen, Germany, Editor D.~Brandt, CERN, 2006. 
\bibitem{HelmutWiedemann_PAP} Helmut Wiedemann, Particle Accelerator Physics, Chapter 5, Fourth edition, Springer 2015, Berlin, ISBN 978-3-319-18316-9. 
\bibitem{petrillo2015} V.~Petrillo {\it et al.}, Polarization of x-gamma radiation produced by a Thomson and Compton inverse scattering, Phys. Rev. ST Accel. Beams {\bf 18} (2015) 110701. 
\bibitem{McMaster1} William H. McMaster, Polarization and the Stokes Parameters, American Journal of Physics {\bf 22}, 351--362 (1954).
\bibitem{McMaster2} William H. McMaster, Matrix Representation of Polarization, Reviews of Modern Physics {\bf 33}, 8--28 (1961).
\bibitem{zhijun_chi2020} Zhijun~Chi, Polarization transfer from a laser to x rays via Thomson scattering with relativistic electrons: A dipole radiation perspective, J. Appl. Phys. {\bf 128}, 244904 (2020).
\bibitem{zhijun_chi2022} Zhijun~Chi, X-ray polarization characteristics in the nonlinear Thomson scattering of a laser with relativistic electrons, Nucl. Instrum. Methods Phys. Res. A {\bf 1033}, 166681 (2022).
\bibitem{Landau} L. Landau, E. Lifchitz, V. Berestetski, L. Pitayevski, Th\'eorie Quantique Relativiste - premi\'ere partie, Chapter X, Section 87, \'Editions MIR, Moscou (1972).
\bibitem{takashi_minipix} Takashi Ari-izumi {\it et al.}, in preparation. 
\end{thebibliography}


\end{document}